# Achievable Angles Between two Compressed Sparse Vectors Under Norm/Distance Constraints Imposed by the Restricted Isometry Property: A Plane Geometry Approach

Ling-Hua Chang and Jwo-Yuh Wu

Department of Electrical Engineering,
Institute of Communications Engineering,
National Chiao Tung University,
1001, Ta-Hsueh Road, Hsinchu, Taiwan
Tel: 886-35-712121, ext. 54524; Fax: 886-35-710116
Email: iamjaung@gmail.com, jywu@cc.nctu.edu.tw

*Abstract*-The angle between two compressed sparse vectors subject to the norm/distance constraints imposed by the restricted isometry property (RIP) of the sensing matrix plays a crucial role in the studies of many compressive sensing (CS) problems. Assuming that (i) $\mathbf{u}$ and $\mathbf{v}$ are two sparse vectors with $\angle(\mathbf{u}, \mathbf{v}) = \theta$ and (ii) the sensing matrix $\Phi$ satisfies RIP, this paper is aimed at analytically characterizing the achievable angles between $\Phi\mathbf{u}$ and $\Phi\mathbf{v}$. Motivated by geometric interpretations of RIP and with the aid of the well-known law of cosines, we propose a plane geometry based formulation for the study of the considered problem. It is shown that all the RIP-induced norm/distance constraints on $\Phi\mathbf{u}$ and $\Phi\mathbf{v}$ can be jointly depicted via a simple geometric diagram in the two-dimensional plane. This allows for a joint analysis of all the considered algebraic constraints from a geometric perspective. By conducting plane geometry analyses based on the constructed diagram, closed-form formulae for the maximal and minimal achievable angles are derived. Computer simulations confirm that the proposed solution is tighter than an existing algebraic-based estimate derived using the polarization identity. The obtained results are used to derive a tighter restricted isometry constant of structured sensing matrices of a certain kind, to wit, those in the form of a product of an orthogonal projection matrix and a random sensing matrix. Follow-up applications to three CS problems, namely, compressed-domain interference cancellation, RIP-based analysis of the orthogonal matching pursuit algorithm, and the study of democratic nature of random sensing matrices are investigated.

*Index Terms:* Compressive sensing; restricted isometry property (RIP); restricted isometry constant (RIC); interference cancellation; orthogonal matching pursuit; plane geometry.

*Suggested Editorial Area:* Signal Processing.

## I. INTRODUCTION

Compressive sensing (CS) is a new technique which exploits sparsity inherent in wide classes of real-world signals so as to facilitate efficient data acquisition, storage, and processing [1-3]. Applications of CS have been found in various areas, including analog-to-digital converters [4], magnetic resonance imaging [5], wireless communications [6-11], sensor networks [11-13], and linear control [14-16], to name just a few. A CS system is typically described by an underdetermined linear equation set

This work is sponsored by the National Science Council of Taiwan under grants NSC 100-2221-E-009-104-MY3 and NSC 100-2628-E-009-008, by the Ministry of Education of Taiwan under the MoE ATU Program, and by the Telecommunication Laboratories, Chunghwa Telecom Co., Ltd. under grant TL-99-G107. Part of the materials of this paper was submitted to *2012 American Control Conference (ACC)*, and *IEEE Wireless Communications and Networking Conference 2012 (IEEE WCNC 2012)*; see [27] and [28], respectively. While [27] is under review, [28] has been accepted for presentation at the time this full journal version is submitted.

$$\mathbf{y} = \mathbf{\Phi s},\qquad(1.1)$$

where $\mathbf{s} \in \mathbb{R}^p$ is a $K$-sparse signal vector ($K \ll p$), $\mathbf{y} \in \mathbb{R}^m$ is the compressed measurement vector, and $\mathbf{\Phi} \in \mathbb{R}^{m\times p}$ is the sensing matrix with $m < p$. To guarantee unique signal identification based on the compressed data, it is required that the sensing matrix $\mathbf{\Phi}$ must satisfy the so-called restricted isometry property (RIP) of order $K$ [17-18], that is, there exists $0 < \delta < 1$ such that

$$(1-\delta)\|\mathbf{s}\|_2^2 \leq \|\mathbf{\Phi s}\|_2^2 \leq (1+\delta)\|\mathbf{s}\|_2^2 \qquad(1.2)$$

holds for all $K$-sparse $\mathbf{s}$. In particular, for two sparse vectors $\mathbf{u}$ and $\mathbf{v}$ supported on $T_u$ and $T_v$ such that $|T_u \cup T_v|$, the cardinality of $T_u \cup T_v$, satisfies $|T_u \cup T_v| \leq K$, it follows from (1.2) that

$$(1-\delta)\|\mathbf{u}-\mathbf{v}\|_2^2 \leq \|\mathbf{\Phi}(\mathbf{u}-\mathbf{v})\|_2^2 \leq (1+\delta)\|\mathbf{u}-\mathbf{v}\|_2^2.\qquad(1.3)$$

Roughly speaking, if a sensing matrix $\mathbf{\Phi}$ satisfies RIP (1.2) with a small restricted isometry constant (RIC) $\delta$, the information about $\mathbf{x}$ remains largely intact upon compression. In addition, (1.3) ensures that the Euclidean distance between two sparse vectors is approximately preserved in the compressed domain; this will guarantee robustness of signal recovery against noise perturbation. Both (1.2) and (1.3) characterize signal identifiability in terms of the norm, or Euclidean distance, of compressed sparse vectors. In the literature of CS, the angle $\angle(\mathbf{\Phi u},\mathbf{\Phi v})$ between a compressed vector pair $\{\mathbf{\Phi u},\mathbf{\Phi v}\}$ plays an important role in many studies regarding stability analyses and performance evaluations. For example, in parameter estimation with compressed measurements [11], $\angle(\mathbf{\Phi u},\mathbf{\Phi v})$ is relevant to the assessment of the estimation errors. Also, in compressed-domain interference cancellation [9-11], in RIP-based analyses of the orthogonal matching pursuit (OMP) algorithm [19], and in the study of the democratic nature of random sensing matrices [21-23], the absolute value of $\cos(\angle(\mathbf{\Phi u},\mathbf{\Phi v}))$ is essential for evaluating the RIC of a structured sensing matrix which characterizes the effective data acquisition process. In the aforementioned works, the analysis resorted to certain upper bounds of $|\cos(\angle(\mathbf{\Phi u},\mathbf{\Phi v}))|$ that are derived by using inequalities such as (1.2) and (1.3) in conjunction with the polarization identity[1]. The bounds obtained via such an algebraic-based RIP analysis, however, are the worst-case estimates [24], and will lead to a pessimistic judgment about the true system performance. Toward more accurate performance evaluations, a fundamental approach is to explicitly specifying the achievable $\angle(\mathbf{\Phi u},\mathbf{\Phi v})$ subject to the norm/distance constraints induced by RIP. In-depth studies of such problems, however, have not been seen in the literature yet.

This paper investigates the maximal and minimal achievable angles between two compressed sparse vectors under norm/distance constraints imposed by RIP. To be more precise, we consider two sparse vectors $\mathbf{u} \in \mathbb{R}^p$ and $\mathbf{v} \in \mathbb{R}^p$ whose supports $T_u$ and $T_v$ satisfy $|T_u \cup T_v| \leq K$. Suppose that the angle between $\mathbf{u}$ and $\mathbf{v}$ is $\angle(\mathbf{u}, \mathbf{v}) = \theta$, that is,

$$\frac{\langle \mathbf{u}, \mathbf{v} \rangle}{\|\mathbf{u}\|_2 \|\mathbf{v}\|_2} = \cos\theta, \tag{1.4}$$

where $0 < \theta \leq \pi/2$ is assumed without loss of generality[2]. In the compressed domain, the angle between $\mathbf{\Phi u}$ and $\mathbf{\Phi v}$ is $\angle(\mathbf{\Phi u}, \mathbf{\Phi v}) = \alpha$, i.e.,

$$\frac{\langle \mathbf{\Phi u}, \mathbf{\Phi v} \rangle}{\|\mathbf{\Phi u}\|_2 \|\mathbf{\Phi v}\|_2} = \cos\alpha. \tag{1.5}$$

---

1. For $\mathbf{u},\mathbf{v} \in \mathbb{R}^p$, the inner product between $\mathbf{u}$ and $\mathbf{v}$ can be expressed as $<\mathbf{u},\mathbf{v}> = \{\|\mathbf{u}+\mathbf{v}\|_2^2 - \|\mathbf{u}-\mathbf{v}\|_2^2\}/4$ [23].

Under RIP (1.2) and (1.3), all we know about $\mathbf{\Phi u}$ and $\mathbf{\Phi v}$ are the plausible values of the respective norms and distance. This implies that the measure of $\alpha$ (or the value of $\cos\alpha$) will lie within a certain range. Given a fixed $\angle(\mathbf{u}, \mathbf{v}) = \theta$, we propose a method for identifying the maximal and the minimal $\alpha$, hereafter denoted respectively by[3] $\alpha_{\max}$ and $\alpha_{\min}$, subject to the RIP-induced norm/distance constraints. Specific technical contributions of this paper can be summarized as follows.

1. By exploiting geometric interpretations of RIP and the well-known law of cosines [25-26], it is shown that the angle between a feasible pair $\{\mathbf{\Phi u}, \mathbf{\Phi v}\}$ has the same measure as the angle determined by one vertex of an auxiliary triangle in the *two-dimensional plane*. This leads to a natural problem formulation on the basis of the plane geometry framework for the study of the achievable $\alpha$. With the aid of the proposed formulation, there is a simple way to construct a geometric diagram depicting all the auxiliary triangles associated with all feasible $\{\mathbf{\Phi u}, \mathbf{\Phi v}\}$. The problem then boils down to searching into the diagram for the two triangles whose corresponding vertices yield, respectively, the largest and smallest angles.

2. The distinctive features of the proposed *plane geometry* based formulation are twofold. Firstly, all the RIP-induced algebraic norm/distance constraints that are relevant to the characterization of the angle $\angle(\mathbf{\Phi u}, \mathbf{\Phi v})$ can be jointly elucidated from a plane geometry perspective. This facilitates a *joint* analysis of all the considered algebraic constraints in the plane geometry setting for the identification of $\alpha_{\max}$ and $\alpha_{\min}$. Such an approach is in marked contrast with the existing algebraic-based method [9], [19], [21], wherein the extreme values of the norm/distance specified by the inequalities (1.2) and (1.3) are employed to obtain mere the worst-case estimate of $\alpha$ (or $\cos\alpha$). Secondly, by conducting plane geometry analyses based on the constructed geometric diagram, *closed-form* formulae of $\alpha_{\max}$ and $\alpha_{\min}$ can be derived, and are then validated through computer simulations.

---

2. If the angle $\angle(\mathbf{u}, \mathbf{v}) = \theta > \pi/2$, then consider instead, say, $\mathbf{u}$ and $-\mathbf{v}$ with $\angle(\mathbf{u}, -\mathbf{v}) = (\pi - \theta) < \pi/2$. Since $\angle(\mathbf{\Phi u}, \mathbf{\Phi v})$ and $\angle(\mathbf{\Phi u}, -\mathbf{\Phi v})$ are supplementary, i.e., $\angle(\mathbf{\Phi u}, \mathbf{\Phi v}) + \angle(\mathbf{\Phi u}, -\mathbf{\Phi v}) = \pi$, $\alpha_{\max}$ and $\alpha_{\min}$ can be directly obtained as the supplements of, respectively, the smallest and largest achievable $\angle(\mathbf{\Phi u}, -\mathbf{\Phi v})$.

3. Unless otherwise specified, the dependence of $\alpha_{\max}$ and $\alpha_{\min}$ on $\theta$ is omitted in the sequel to conserve notation.

3. Applications of the obtained analytic results to CS are discussed. First of all, the achievable RIC of a structured sensing matrix of the form $\mathbf{P\Phi}$, where $\mathbf{P}$ is a certain orthogonal projection matrix and $\mathbf{\Phi}$ is a random sensing matrix, is investigated. Matrices of this kind were found in, e.g., compressed-domain interference cancellation [9-11], RIP-based analyses of the OMP algorithm [19], and characterization of the democratic nature of random sensing matrices [21-22]. Based on the knowledge of $\alpha_{\max}$ and $\alpha_{\min}$, we derive a closed-form formula for the RIC of $\mathbf{P\Phi}$. Our solution is shown to be tighter than an existing estimate proposed in [9], [11], [19]. The impacts of this result on the three CS problems mentioned above are then discussed.

*i) Compressed-Domain Interference Cancellation:* In this problem, $\mathbf{P\Phi}$ represents the effective sensing system matrix after the interference is removed via orthogonal projection [9]. Thus, the RIC of $\mathbf{P\Phi}$ plays a decisive role in performance evaluation. The established result, namely, $\mathbf{P\Phi}$ enjoys a tighter RIC, can lead to a tighter estimate of the signal reconstruction error. In addition, to achieve a target threshold of the RIC of $\mathbf{P\Phi}$, our result allows for a less restricted requirement on the RIC of the original sensing matrix $\mathbf{\Phi}$ (i.e., the RIC of $\mathbf{\Phi}$ is allowed to be larger). Since the required number of measurements decreases with the RIC of $\mathbf{\Phi}$ [3], this in turn implies that the measurement size can be further reduced.

*ii) RIP-Based Analysis of OMP Algorithm:* We derive a new sufficient condition under which the OMP algorithm can perfectly recover a *K*-sparse vector in exactly *K* iterations. Compared with the recently reported result in [20], our solution provides a less restricted upper bound on the required RIC of $\mathbf{\Phi}$. Hence, the OMP algorithm can achieve exact signal recovery with a reduced measurement size. Our analytic study is further confirmed by computer simulations.

*iii) Democratic Nature of Random Sensing Matrices:* Given a random sensing matrix $\mathbf{\Phi}$, let $\tilde{\mathbf{\Phi}}$ be a sub-matrix obtained by removing from $\mathbf{\Phi}$ a small and randomly chosen subset of rows. The RIC of $\tilde{\mathbf{\Phi}}$ is crucial for characterizing the so-called "democratic property" of random sensing matrices [21]. We show that the RIC of $\tilde{\mathbf{\Phi}}$ is smaller than an existing estimate: this implies that random sensing indeed provides better robustness against the loss of measurements.

The rest of this paper is organized as follows. Section II presents the proposed plane geometry based formulation. The maximal and minimal achievable angles are derived, respectively, in Sections III and IV. Section V discusses the connection between the presented analytic study with previous works; computer simulations are also conducted to evidence the proposed analytic solutions. Section VI investigates the applications of the obtained results to CS. Section VII is the conclusion. To ease reading, all supporting technical proofs of the established mathematical lemmas and theorems are relegated to the appendix.

# II. PROPOSED PLANE GEOMETRY BASED FORMULATION

This section introduces the proposed plane geometry based analysis framework for identifying $\alpha_{\max}$ and $\alpha_{\min}$. After highlighting some intuitive motivations, Section II-A presents the proposed problem formulation. Section II-B constructs a geometric diagram in the two-dimensional plane that shows concrete geometric depictions of all the feasible $\angle(\mathbf{\Phi u}, \mathbf{\Phi v})$ under RIP. Section II-C provides more explicit characterizations about $\alpha_{\max}$ and $\alpha_{\min}$ within the proposed plane geometry framework. The derivations of $\alpha_{\max}$ and $\alpha_{\min}$ will be given in Sections III and IV.

Recall that, under RIP (1.2) and (1.3), all we know about $\mathbf{\Phi u}$ and $\mathbf{\Phi v}$ is the maximal, and minimal, achievable norms and distance. To identify $\alpha_{\max}$ and $\alpha_{\min}$, the first step is to find a mathematical relation which can delineate the connection between the norm/distance information about the pair $\{\mathbf{\Phi u}, \mathbf{\Phi v}\}$ and the resultant angle $\angle(\mathbf{\Phi u}, \mathbf{\Phi v}) = \alpha$. Hopefully, such a formula can moreover provide distinctive insights to facilitate an analytic study of the considered problem from a geometric perspective. The ideas above can be realized by the well-known *law of cosines*, which states that the angle $\angle(\mathbf{\Phi u}, \mathbf{\Phi v}) = \alpha$ can be determined by $\|\mathbf{\Phi u}\|_2$, $\|\mathbf{\Phi v}\|_2$, and $\|\mathbf{\Phi (u-v)}\|_2$ as

$$\cos \alpha = \frac{\|\mathbf{\Phi u}\|_2^2 + \|\mathbf{\Phi v}\|_2^2 - \|\mathbf{\Phi (u-v)}\|_2^2}{2\|\mathbf{\Phi u}\|_2 \|\mathbf{\Phi v}\|_2}. \tag{2.1}$$

Hence, at least conceptually, by taking account of all the available knowledge about $\|\mathbf{\Phi u}\|_2$, $\|\mathbf{\Phi v}\|_2$, and $\|\mathbf{\Phi (u-v)}\|_2$ under RIP, it is possible to characterize the achievable $\alpha$ and, in particular, to identify $\alpha_{\max}$ and $\alpha_{\min}$. The unique advantage of (2.1) is its underlying geometric interpretation: we can think of $\|\mathbf{\Phi u}\|_2$, $\|\mathbf{\Phi v}\|_2$, and $\|\mathbf{\Phi (u-v)}\|_2$ as the three sides of a triangle in $\mathbb{R}^m$, and $\angle(\mathbf{\Phi u}, \mathbf{\Phi v})$ has the same measure as the angle defined by a certain vertex of this triangle. As will be shown later, such a simple and concrete geometric view of $\angle(\mathbf{\Phi u}, \mathbf{\Phi v})$ will allow for a natural problem formation within the plane geometry framework. We shall note that the angle between two vectors is invariant to scaling of the respective norms, that is, if $\mathbf{u}$ and $\mathbf{v}$ are such that $\angle(\mathbf{u}, \mathbf{v}) = \theta$ and $\angle(\mathbf{\Phi u}, \mathbf{\Phi v}) = \alpha$, then it follows $\angle(c_1\mathbf{u}, c_2\mathbf{v}) = \theta$ and $\angle(c_1\mathbf{\Phi u}, c_2\mathbf{\Phi v}) = \alpha$ regardless of any positive scalars $c_1$ and $c_2$ (see (1.4) and (1.5)). Hence, we assume in the sequel that $\|\mathbf{u}\|_2^2 = \|\mathbf{v}\|_2^2 = 1$.

Before introducing the proposed formulation, let us first specify the relevant norm/distance information about $\mathbf{\Phi u}$ and $\mathbf{\Phi v}$ that is revealed by RIP. The plausible values of $\|\mathbf{\Phi u}\|_2^2$ and $\|\mathbf{\Phi v}\|_2^2$ under RIP (1.2) are

$$1 - \delta \leq \|\mathbf{\Phi u}\|_2^2 \leq 1 + \delta \ \text{ and } \ 1 - \delta \leq \|\mathbf{\Phi v}\|_2^2 \leq 1 + \delta. \tag{2.2}$$

Also, the achievable distance between $\mathbf{\Phi u}$ and $\mathbf{\Phi v}$ can be directly obtained from (1.3) and the law of cosines as

$$d_{\min}^2 \leq \|\mathbf{\Phi(u - v)}\|_2^2 \leq d_{\max}^2, \tag{2.3}$$

in which

$$d_{\max}^2 \triangleq (1+\delta)\|\mathbf{u} - \mathbf{v}\|_2^2 = (1+\delta) \cdot \{\|\mathbf{u}\|_2^2 + \|\mathbf{v}\|_2^2 - 2\|\mathbf{u}\|_2\|\mathbf{v}\|_2 \cos\theta\} = 2(1+\delta)(1-\cos\theta), \tag{2.4}$$

and

$$d_{\min}^2 \triangleq (1-\delta)\|\mathbf{u} - \mathbf{v}\|_2^2 = (1-\delta) \cdot \{\|\mathbf{u}\|_2^2 + \|\mathbf{v}\|_2^2 - 2\|\mathbf{u}\|_2\|\mathbf{v}\|_2 \cos\theta\} = 2(1-\delta)(1-\cos\theta). \tag{2.5}$$

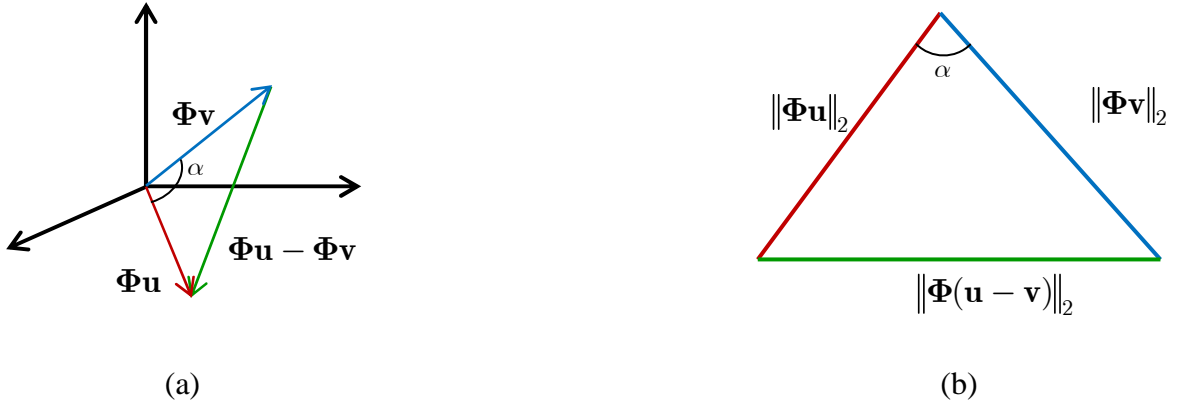

Figure 1. (a) The triangle defined by three $m$-dimensional vectors $\mathbf{\Phi u}$, $\mathbf{\Phi v}$, and $\mathbf{\Phi(u - v)}$ (we use $m = 3$ for illustration). (b) Depiction of the auxiliary triangle in the two-dimensional plane; in particular, based on the law of cosines (2.1), the angle $\alpha$ between $\mathbf{\Phi u}$ and $\mathbf{\Phi v}$ can be computed using $\|\mathbf{\Phi u}\|_2$, $\|\mathbf{\Phi v}\|_2$, and $\|\mathbf{\Phi(u-v)}\|_2$.

We note that the law of cosines (2.1) can also be expressed as

$$\cos\alpha = -\cos(\pi - \alpha) = -\left(\frac{\|\mathbf{\Phi u}\|_2^2 + \|\mathbf{\Phi v}\|_2^2 - \|\mathbf{\Phi(u + v)}\|_2^2}{2\|\mathbf{\Phi u}\|_2 \|\mathbf{\Phi v}\|_2}\right). \tag{2.6}$$

Hence, in addition to $\|\mathbf{\Phi u}\|_2^2$, $\|\mathbf{\Phi v}\|_2^2$, and $\|\mathbf{\Phi(u - v)}\|_2^2$, equation (2.6) implies that the knowledge of $\|\mathbf{\Phi(u + v)}\|_2^2$ is also needed for characterizing $\cos\alpha$. As a result, the following constraint on $\|\mathbf{\Phi(u + v)}\|_2^2$ imposed by RIP should also be taken into account:

$$\tilde{d}_{\min}^2 \leq \|\mathbf{\Phi(u + v)}\|_2^2 \leq \tilde{d}_{\max}^2, \tag{2.7}$$

where

$$\tilde{d}_{\max}^2 \triangleq (1+\delta)\|\mathbf{u} + \mathbf{v}\|_2^2 = (1+\delta) \cdot \{\|\mathbf{u}\|_2^2 + \|\mathbf{v}\|_2^2 + 2\|\mathbf{u}\|_2\|\mathbf{v}\|_2 \cos\theta\} = 2(1+\delta)(1+\cos\theta), \tag{2.8}$$

and

$$\tilde{d}_{\min}^2 \triangleq (1-\delta)\|\mathbf{u} + \mathbf{v}\|_2^2 = (1-\delta) \cdot \{\|\mathbf{u}\|_2^2 + \|\mathbf{v}\|_2^2 + 2\|\mathbf{u}\|_2\|\mathbf{v}\|_2 \cos\theta\} = 2(1-\delta)(1+\cos\theta). \tag{2.9}$$

The conditions (2.2), (2.3), and (2.7) can be regarded as the full characterization of the norm/distance information about $\mathbf{\Phi u}$ and $\mathbf{\Phi v}$ under RIP. Our task is to find $\alpha_{\max}$ and $\alpha_{\min}$ based solely on (2.2), (2.3), and (2.7). The following notation will be used throughout the rest of this paper.

$\overline{PQ}$: the line segment between the two points $P$ and $Q$ in the two-dimensional plane.

$|\overline{PQ}|$: length of the line segment $\overline{PQ}$.

$\triangle ABC$: the triangle with three vertexes denoted by $A$, $B$, and $C$.

$\angle ABC$: the angle with $\overline{BA}$ and $\overline{BC}$ as sides and $B$ as the vertex.

$\mathcal{C}(O,r)$: the circle centered at $O$ with radius $r$, thus $\mathcal{C}(O,r) = \{P \in \mathbb{R}^2 \mid \|P - O\|_2 = r\}$.

$\mathcal{CV}(P_1, P_2)$: the connected curve in the two-dimensional plane with $P_1$ and $P_2$ as the two end points.

*A. Problem Formulation via Plane Geometry*

The proposed approach is motivated by the crucial observation: the triangle in the $m$-dimensional Euclidean space with three sides given by $\|\mathbf{\Phi u}\|_2$, $\|\mathbf{\Phi v}\|_2$, and $\|\mathbf{\Phi(u-v)}\|_2$ can be directly depicted in the *two dimensional plane* (see Figure 1). As a result, the study of the achievable angles between the two $m$-dimensional compressed vectors $\mathbf{\Phi u}$ and $\mathbf{\Phi v}$ can be formulated by means of such an auxiliary triangle from a plane geometry setting. In this way, the "algebraic" constraints (2.2), (2.3), and (2.7) can be jointly elucidated from a geometric perspective; more importantly, a joint analysis of these constraints within the plane geometry framework can therefore be carried out to derive closed-form solutions for $\alpha_{\max}$ and $\alpha_{\min}$, as will be seen later.

To begin with, we shall leverage the idea of auxiliary triangles to translate the algebraic constraints (2.2), (2.3), and (2.7) into concrete geometric depictions in the two-dimensional plane. Toward this end, let us first pick a plausible compressed distance $\|\mathbf{\Phi(u-v)}\|_2 = d$, $d_{\min} \leq d \leq d_{\max}$, and construct all the feasible auxiliary triangles with a common bottom of length $d$. This can be done by constructing a line segment of length $d$, and then going on to find the locations of all feasible top vertices, whereof each one has the property that (i) the lengths of the two sides (equal to $\|\mathbf{\Phi u}\|_2$ and $\|\mathbf{\Phi v}\|_2$, respectively) obey the norm constraint (2.2), and (ii) the resultant $\|\mathbf{\Phi(u+v)}\|_2^2$ satisfies (2.7).

Specifically, for a plausible $d$, let us construct a segment, say $\overline{B_d C_d}$, in the plane with $|\overline{B_d C_d}| = \|\mathbf{\Phi(u-v)}\|_2 = d$ (see Figure 2-(a), shown on page 9). We can first determine the region $\widehat{\mathbf{\Omega}}(d)$ such that, if $D_d \in \widehat{\mathbf{\Omega}}(d)$, then $|\overline{D_d B_d}| = \|\mathbf{\Phi u}\|_2$ and $|\overline{D_d C_d}| = \|\mathbf{\Phi v}\|_2$ fulfill (2.2), i.e., $\sqrt{1-\delta} \leq |\overline{D_d B_d}| \leq \sqrt{1+\delta}$ and $\sqrt{1-\delta} \leq |\overline{D_d C_d}| \leq \sqrt{1+\delta}$. For this, let us construct four circles, i.e., $\mathcal{C}(B_d, \sqrt{1+\delta})$, $\mathcal{C}(B_d, \sqrt{1-\delta})$, $\mathcal{C}(C_d, \sqrt{1+\delta})$, and $\mathcal{C}(C_d, \sqrt{1-\delta})$. Clearly, $\widehat{\mathbf{\Omega}}(d)$ is simply the intersection of the two annular regions defined by the four circles[4]; see the two grey regions in Figure 2-(a), and due to symmetry it suffices to consider the one in the upper half. For a fixed $D_d \in \widehat{\mathbf{\Omega}}(d)$, the triangle $\triangle D_d B_d C_d$ is thus associated with the magnitude triple $(\|\mathbf{\Phi u}\|_2, \|\mathbf{\Phi v}\|_2, \|\mathbf{\Phi(u-v)}\|_2) = (|\overline{D_d B_d}|, |\overline{D_d C_d}|, |\overline{B_d C_d}| = d)$, and the angle $\alpha$ between the $m$-dimensional vector pair $\{\mathbf{\Phi u}, \mathbf{\Phi v}\}$ is

---

4. The two annular regions defined by the four circles must overlap since, by invoking the definition of $d_{\max}$ in (2.4), it is easy to show that the sum of the radii of the two outer circles is no less than $d$, that is, $\sqrt{1+\delta} + \sqrt{1+\delta} = 2\sqrt{1+\delta} \geq d_{\max} \geq d$.

exactly given by $\alpha = \angle B_d D_d C_d$. Nevertheless, not every $D_d \in \widehat{\Omega}(d)$ is feasible, since the corresponding $\|\mathbf{\Phi}(\mathbf{u}+\mathbf{v})\|_2^2$ may fail to satisfy the constraint (2.7). To further take (2.7) into account for identifying the locations of all feasible top vertices, we shall rewrite (2.7) in an alternative, yet equivalent, form more amenable to handle. For this we first note that

$$\cos\alpha = \frac{\langle \mathbf{\Phi}\mathbf{u}, \mathbf{\Phi}\mathbf{v}\rangle}{\|\mathbf{\Phi}\mathbf{u}\|_2 \|\mathbf{\Phi}\mathbf{v}\|_2} \stackrel{(a)}{=} \frac{\|\mathbf{\Phi}\mathbf{u}+\mathbf{\Phi}\mathbf{v}\|_2^2 - \|\mathbf{\Phi}\mathbf{u}-\mathbf{\Phi}\mathbf{v}\|_2^2}{4\|\mathbf{\Phi}\mathbf{u}\|_2 \|\mathbf{\Phi}\mathbf{v}\|_2} \stackrel{(b)}{=} \frac{\|\mathbf{\Phi}\mathbf{u}+\mathbf{\Phi}\mathbf{v}\|_2^2 - d^2}{4\|\mathbf{\Phi}\mathbf{u}\|_2 \|\mathbf{\Phi}\mathbf{v}\|_2}, \tag{2.10}$$

where (a) follows from the polarization identity, and (b) holds since $\|\mathbf{\Phi}(\mathbf{u}-\mathbf{v})\|_2 = d$. Equation (2.10) then implies

$$\|\mathbf{\Phi}\mathbf{u}+\mathbf{\Phi}\mathbf{v}\|_2^2 = 4\|\mathbf{\Phi}\mathbf{u}\|_2 \|\mathbf{\Phi}\mathbf{v}\|_2 \cos\alpha + d^2. \tag{2.11}$$

By again invoking the law of cosines (2.1) and $\|\mathbf{\Phi}(\mathbf{u}-\mathbf{v})\|_2 = d$, we can rewrite (2.11) as

$$\|\mathbf{\Phi}\mathbf{u}+\mathbf{\Phi}\mathbf{v}\|_2^2 = 2(\|\mathbf{\Phi}\mathbf{u}\|_2^2 + \|\mathbf{\Phi}\mathbf{v}\|_2^2) - d^2 = 2(|\overline{D_d B_d}|^2 + |\overline{D_d C_d}|^2) - d^2, \tag{2.12}$$

where the last equality holds since in our setting $\|\mathbf{\Phi}\mathbf{u}\|_2 = |\overline{D_d B_d}|$ and $\|\mathbf{\Phi}\mathbf{v}\|_2 = |\overline{D_d C_d}|$. Based on (2.12) together with some straightforward manipulations, the constraint (2.7) on $\|\mathbf{\Phi}(\mathbf{u}+\mathbf{v})\|_2^2$ can be equivalently rewritten in terms of $|\overline{D_d B_d}|^2 + |\overline{D_d C_d}|^2$ as follows:

$$\frac{\tilde{d}_{\min}^2 + d^2}{2} \leq |\overline{D_d B_d}|^2 + |\overline{D_d C_d}|^2 \leq \frac{\tilde{d}_{\max}^2 + d^2}{2}. \tag{2.13}$$

Hence, the *feasible top-vertex set* associated with $\|\mathbf{\Phi}(\mathbf{u}-\mathbf{v})\|_2 = |\overline{B_d C_d}| = d$, hereafter denoted by $\mathbf{\Omega}(d)$, consists of all $D_d \in \widehat{\mathbf{\Omega}}(d)$ that also satisfy (2.13), namely,

$$\mathbf{\Omega}(d) = \left\{ D_d \mid D_d \in \widehat{\mathbf{\Omega}}(d), \frac{\tilde{d}_{\min}^2 + d^2}{2} \leq |\overline{D_d B_d}|^2 + |\overline{D_d C_d}|^2 \leq \frac{\tilde{d}_{\max}^2 + d^2}{2} \right\}. \tag{2.14}$$

From (2.14), it is easy to see that $\mathbf{\Omega}(d)$ can be obtained by removing from $\widehat{\mathbf{\Omega}}(d)$ the following two parts:

$$\widehat{\mathbf{\Omega}}_t(d) = \left\{ D_d \mid D_d \in \widehat{\mathbf{\Omega}}(d), |\overline{D_d B_d}|^2 + |\overline{D_d C_d}|^2 > \frac{\tilde{d}_{\max}^2 + d^2}{2} \right\}, \tag{2.15}$$

and

$$\widehat{\mathbf{\Omega}}_b(d) = \left\{ D_d \mid D_d \in \widehat{\mathbf{\Omega}}(d), |\overline{D_d B_d}|^2 + |\overline{D_d C_d}|^2 < \frac{\tilde{d}_{\min}^2 + d^2}{2} \right\}. \tag{2.16}$$

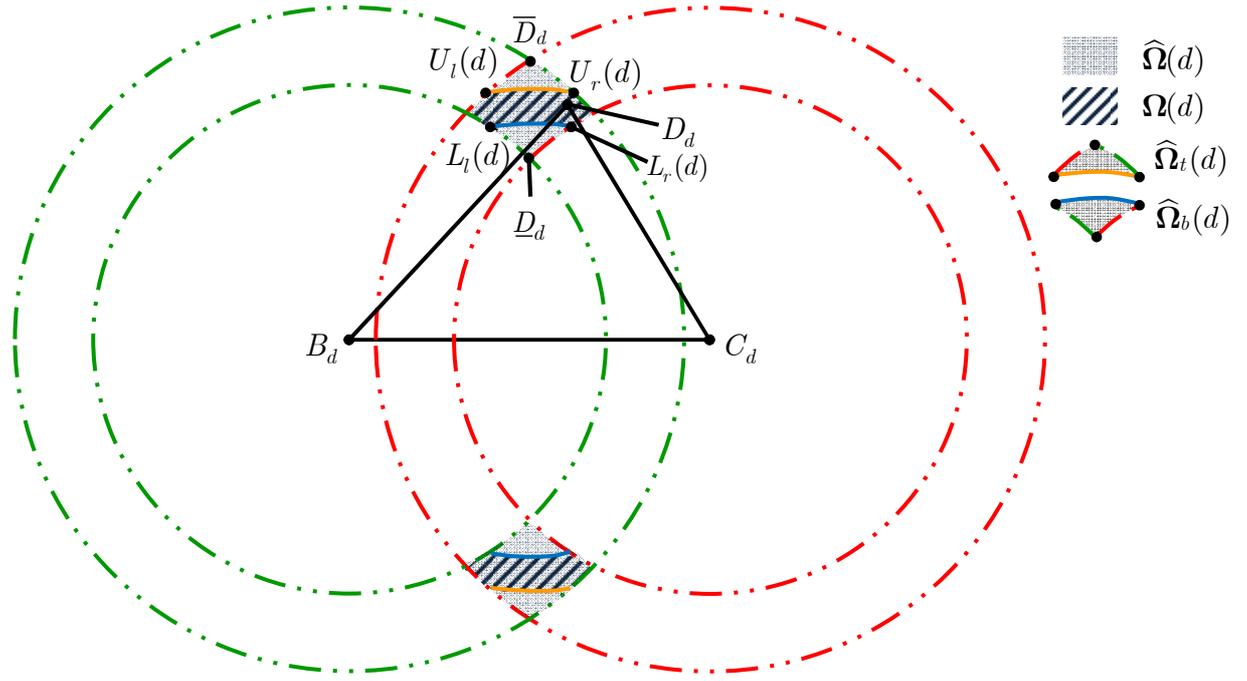

(a)

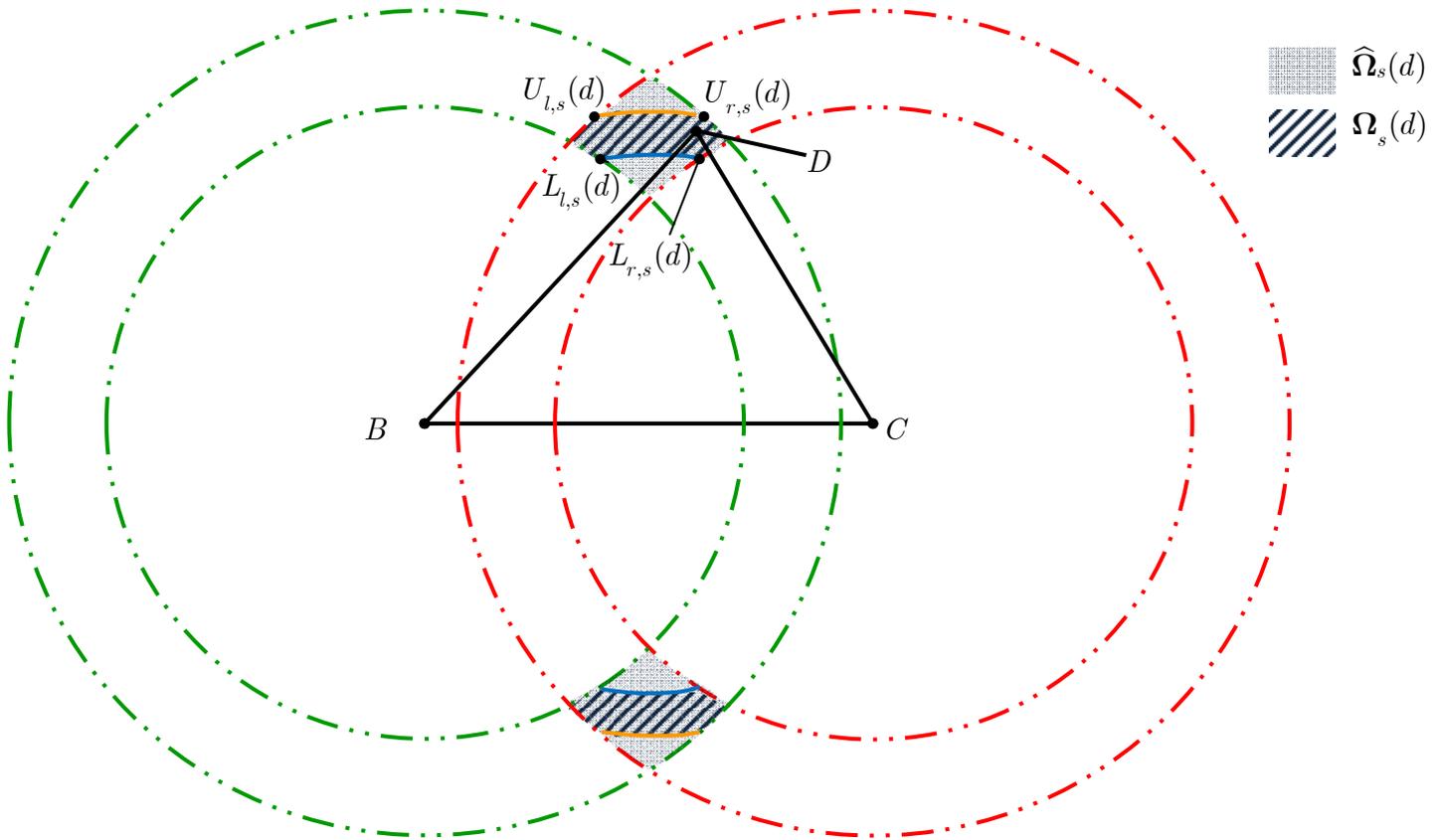

(b)

Figure 2. (a) Schematic description of the feasible top-vertex set $\mathbf{\Omega}(d)$ associated with $\|\mathbf{\Phi}(\mathbf{u}-\mathbf{v})\|_2 = |\overline{B_d C_d}| = d$; (b) The dilated $\mathbf{\Omega}_s(d)$ is similar to $\mathbf{\Omega}(d)$. The triangle $\triangle D_d B_d C_d$ in Figure 2-(a) is similar to $\triangle DBC$ in Figure 2-(b), thereby $\angle D_d B_d C_d = \angle DBC$.

Notably, $\widehat{\boldsymbol{\Omega}}_t(d)$ and $\widehat{\boldsymbol{\Omega}}_b(d)$, both in general nonempty[5], represent the top and bottom "corner regions" of $\widehat{\boldsymbol{\Omega}}(d)$, respectively. The schematic depiction of $\boldsymbol{\Omega}(d)$ is therefore shown as the black dashed region in Figure 2-(a); the top and bottom edges of $\boldsymbol{\Omega}(d)$ are characterized by, respectively, the two curves

$$\mathcal{CV}(U_l(d), U_r(d)) = \{D_d \mid D_d \in \boldsymbol{\Omega}(d), |\overline{D_d B_d}|^2 + |\overline{D_d C_d}|^2 = (\tilde{d}_{\max}^2 + d^2)/2\} \quad (2.17)$$

and

$$\mathcal{CV}(L_l(d), L_r(d)) = \{D_d \mid D_d \in \boldsymbol{\Omega}(d), |\overline{D_d B_d}|^2 + |\overline{D_d C_d}|^2 = (\tilde{d}_{\min}^2 + d^2)/2\}. \quad (2.18)$$

Associated with each $\|\boldsymbol{\Phi}(\mathbf{u} - \mathbf{v})\|_2 = d$ fulfilling (2.3) we have been able to specify the feasible top-vertex set $\boldsymbol{\Omega}(d)$. In particular, for $\boldsymbol{\Phi}\mathbf{u}$ and $\boldsymbol{\Phi}\mathbf{v}$ satisfying (2.2) and (2.7) subject to $\|\boldsymbol{\Phi}(\mathbf{u} - \mathbf{v})\|_2 = d$, the angle $\angle(\boldsymbol{\Phi}\mathbf{u}, \boldsymbol{\Phi}\mathbf{v})$ is exactly given by $\angle(\boldsymbol{\Phi}\mathbf{u}, \boldsymbol{\Phi}\mathbf{v}) = \angle B_d D_d C_d$ for some $D_d \in \boldsymbol{\Omega}(d)$; hence, we have

$$\left\{\angle(\boldsymbol{\Phi}\mathbf{u}, \boldsymbol{\Phi}\mathbf{v}) \mid \boldsymbol{\Phi}\mathbf{u} \text{ and } \boldsymbol{\Phi}\mathbf{v} \text{ satisfy (2.2) and (2.7) subject to } \|\boldsymbol{\Phi}(\mathbf{u} - \mathbf{v})\|_2 = d\right\} = \left\{\angle B_d D_d C_d \mid D_d \in \boldsymbol{\Omega}(d)\right\}$$
(2.19)

The significance of such a formulation is that all the considered norm/distance constraints imposed by RIP can be jointly characterized via simple and concrete geometric depictions in the two-dimensional plane. Based on the above plane geometry framework, the proposed approach for the identification of $\alpha_{\max}$ and $\alpha_{\min}$ is shown next.

*B. Proposed Approach via Similarity*

Given an $\boldsymbol{\Omega}(d)$ constructed as above, one may immediately proceed to seek among all $D_d \in \boldsymbol{\Omega}(d)$ to find

$$\alpha_{\max}(d) = \max_{D_d \in \boldsymbol{\Omega}(d)} \angle B_d D_d C_d \text{ and } \alpha_{\min}(d) = \min_{D_d \in \boldsymbol{\Omega}(d)} \angle B_d D_d C_d$$

associated with the particular $d$. Once $\alpha_{\max}(d)$'s and $\alpha_{\min}(d)$'s for all feasible $d$ are obtained, the maximal and minimal $\alpha$ can be determined as

$$\alpha_{\max} = \max_d \alpha_{\max}(d) \text{ and } \alpha_{\min} = \min_d \alpha_{\min}(d).$$

Such a method, even though conceptually simple, is nonetheless a daunting task since there are un-countably many candidate $d$. This thus motivates us to devise afresh an alternative free of the aforementioned drawbacks. Ideally, if we can come up with a *single* diagram tractably depicting *all* feasible auxiliary triangles $\Delta D_d B_d C_d$ associated with *all* compressed distances $d_{\min} \leq d \leq d_{\max}$, it only remains to consider such a diagram for the identification of $\alpha_{\max}$ and $\alpha_{\min}$. To realize this idea, we resort to the technique of *similarity* in the plane geometry analysis.

---

5. $\widehat{\boldsymbol{\Omega}}_t(d)$ contains the point $\overline{D}_d$, the intersection of the two circles $\mathcal{C}(B_d, \sqrt{1+\delta})$ and $\mathcal{C}(C_d, \sqrt{1+\delta})$, unless $d = d_{\max}$. This is because $|\overline{\overline{D}_d B_d}|^2 + |\overline{\overline{D}_d C_d}|^2 = 2(1+\delta) \stackrel{(a)}{=} (\tilde{d}_{\max}^2 + d_{\max}^2)/2 \stackrel{(b)}{\geq} (\tilde{d}_{\max}^2 + d^2)/2$, where (a) follows from (2.4) and (2.8), and (b) holds since $d_{\max} \geq d$; inequality (b) becomes equality when $d = d_{\max}$. In an analogous way, it can be verified that $\widehat{\boldsymbol{\Omega}}_b(d)$ contains $\underline{D}_d$, the intersection of $\mathcal{C}(B_d, \sqrt{1-\delta})$ and $\mathcal{C}(C_d, \sqrt{1-\delta})$, and is thus nonempty (unless $d = d_{\min}$).

***Definition 2.1 [25]:*** Two figures in the plane are similar whenever one is congruent to a dilation of the other. □

A well-known result is that, if two polygons are similar, the corresponding sides are in proportion, and the corresponding angles must be equal (the so-called conformal property) [26]. Now, suppose we have a figure similar to Figure 2-(a). For each $\Delta D_d B_d C_d$, there is one and only one corresponding triangle, say, $\Delta DBC$, in the similar figure, and $\Delta D_d B_d C_d$ is similar to $\Delta DBC$. Thanks to the conformal property, it follows $\angle B_d D_d C_d = \angle BDC$. Given this fact, we can instead focus on the similar figure as far as the characterization of achievable $\angle B_d D_d C_d$ is concerned. With the aid of judiciously constructed similar figures for all $d_{\min} \leq d \leq d_{\max}$, there is a simple way to obtain a single diagram depicting *all* the similar feasible auxiliary triangles $\Delta DBC$. Therefore, a unified and tractable plane geometry analysis can be conducted to identify $\alpha_{\max}$ and $\alpha_{\min}$.

The construction of such a diagram basically involves two steps as detailed below.

(I) *Construction of the Similar Figure Associated with a Plausible d:* For a plausible compressed distance $\|\mathbf{\Phi}(\mathbf{u} - \mathbf{v})\|_2 = |\overline{B_d C_d}| = d$, let us uniformly stretch all geometric objects in Figure 2-(a) by the scale $d_{\max}/d \, (\geq 1)$ to obtain the dilated similar figure as depicted in Figure 2-(b), in which $\mathbf{\Omega}_s(d)$ is similar to $\mathbf{\Omega}(d)$ and, for each $D_d \in \mathbf{\Omega}(d)$, there exists one and only one corresponding $D \in \mathbf{\Omega}_s(d)$ with which $\Delta DBC$ is similar to $\Delta D_d B_d C_d$. The conformal characteristic of similarity asserts $\angle B_d D_d C_d = \angle BDC$, thus

$$\{\angle B_d D_d C_d \mid D_d \in \mathbf{\Omega}(d)\} = \{\angle BDC \mid D \in \mathbf{\Omega}_s(d)\}. \tag{2.20}$$

Notably, through the proposed dilation procedure the length of the enlarged $\|\mathbf{\Phi}(\mathbf{u} - \mathbf{v})\|_2 = |\overline{B_d C_d}|$ is then *fixed* to be $d \times (d_{\max}/d) = |\overline{BC}| = d_{\max}$ irrespective of *d*. That is to say, *all* the similar auxiliary triangles associated with *all* plausible *d*'s are with a *common bottom of identical length*, equal to $d_{\max}$.

Specific procedures for constructing the similar figure as mentioned above are:

(a) Draw the line segment $\overline{BC}$ with $|\overline{BC}| = d_{\max}$.
(b) Obtain $\widehat{\mathbf{\Omega}}_s(d)$, the dilation of $\widehat{\mathbf{\Omega}}(d)$ in Figure 2-(a), as the intersection of the annular region defined by the four enlarged circles $\mathcal{C}(B, \sqrt{1+\delta} \times (d_{\max}/d))$, $\mathcal{C}(B, \sqrt{1-\delta} \times (d_{\max}/d))$, $\mathcal{C}(C, \sqrt{1+\delta} \times (d_{\max}/d))$, and $\mathcal{C}(C, \sqrt{1-\delta} \times (d_{\max}/d))$.

(c) Determine $\mathbf{\Omega}_s(d)$ by specifying its top and bottom edges, which are the dilations of, respectively, $\mathcal{CV}(U_l(d), U_r(d))$ in (2.17) and $\mathcal{CV}(L_l(d), L_r(d))$ in (2.18). For this we note that, if $\overline{DB}$ and $\overline{DC}$ associated with $D \in \mathbf{\Omega}_s(d)$ are, respectively, the dilated $\overline{D_d B_d}$ and $\overline{D_d C_d}$, where $D_d \in \mathbf{\Omega}(d)$, it immediately follows that $|\overline{DB}|^2 = |\overline{D_d B_d}|^2 \times (d_{\max}/d)^2$ and $|\overline{DC}|^2 = |\overline{D_d C_d}|^2 \times (d_{\max}/d)^2$. With this fact in mind, the dilated top and bottom edges can be accordingly determined from (2.17) and (2.18) as

$$\mathcal{CV}(U_{l,s}(d), U_{r,s}(d)) = \left\{ D \,\middle|\, D \in \mathbf{\Omega}_s(d), |\overline{DB}|^2 + |\overline{DC}|^2 = \frac{\tilde{d}_{\max}^2 + d^2}{2} \times \left(\frac{d_{\max}^2}{d^2}\right) = \underbrace{\frac{\tilde{d}_{\max}^2(d_{\max}^2/d^2) + d_{\max}^2}{2}}_{:=\gamma_t(d)} \right\}$$

(2.21)

and

$$\mathcal{CV}(L_{l,s}(d), L_{r,s}(d)) = \left\{ D \,\middle|\, D \in \mathbf{\Omega}_s(d), |\overline{DB}|^2 + |\overline{DC}|^2 = \frac{\tilde{d}_{\min}^2 + d^2}{2} \times \left(\frac{d_{\max}^2}{d^2}\right) = \underbrace{\frac{\tilde{d}_{\min}^2(d_{\max}^2/d^2) + d_{\max}^2}{2}}_{:=\gamma_b(d)} \right\}.$$

(2.22)

(II) *Depiction of all Similar Feasible Auxiliary Triangles in one Diagram:* With $\overline{BC}$ as the common referenced line segment ($|\overline{BC}| = d_{\max}$), use steps (b) and (c) in part (I) to construct $\mathbf{\Omega}_s(d)$'s for all $d_{\min} \leq d \leq d_{\max}$. Observe that, as $d$ decreases from $d_{\max}$ to $d_{\min}$, the radii of the four circles constructed in step (b) will increase, and so do $\gamma_t(d)$ in (2.21) and $\gamma_b(d)$ in (2.22). Hence, as $d$ decreases, the intersection of the annular region defined by the four circles, as well as the dilated top and bottom edges in (2.21) and (2.22), is "pushed away" from $\overline{BC}$: this implies that $\mathbf{\Omega}_s(d)$ continuously "moves upward" as $d$ decreases from $d_{\max}$ to $d_{\min}$ (see Figure 3 on the next page). The *joint feasible top-vertex set* $\mathbf{\Omega}$ is obtained as the union of all $\mathbf{\Omega}_s(d)$'s, namely,

$$\mathbf{\Omega} = \bigcup_{d_{\min} \leq d \leq d_{\max}} \mathbf{\Omega}_s(d). \tag{2.23}$$

Note that, in the two-dimensional plane, $\mathbf{\Omega}$ is the collection of the overlaid $\mathbf{\Omega}_s(d)$'s, with $\mathbf{\Omega}_s(d_{\min})$ on the top and $\mathbf{\Omega}_s(d_{\max})$ at the bottom. The depiction of $\mathbf{\Omega}$ is shown as the grey part in Figure 3. As a result, each similar feasible auxiliary triangle can be depicted in Figure 3 as $\triangle DBC$ for some $D \in \mathbf{\Omega}$, thereby

$$\bigcup_{d_{\min} \leq d \leq d_{\max}} \{\angle BDC \mid D \in \mathbf{\Omega}_s(d)\} = \{\angle BDC \mid D \in \mathbf{\Omega}\}. \tag{2.24}$$

Now, we can reach the following conclusions:

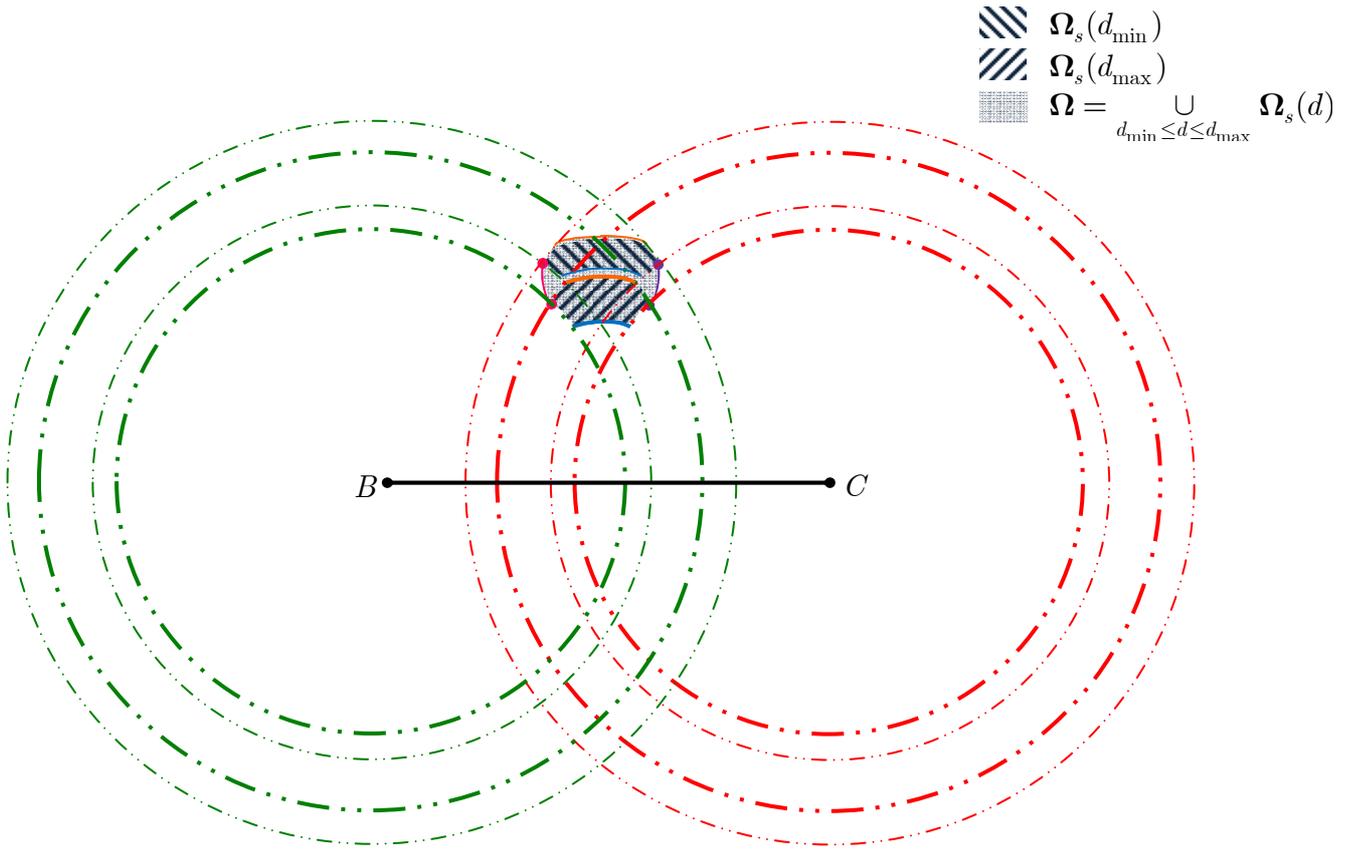

Figure 3. Schematic description of the joint feasible top-vertex set $\mathbf{\Omega}$.

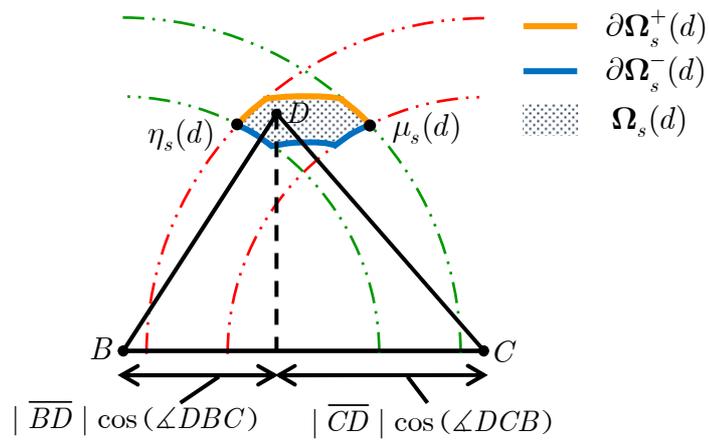

Figure 4. Decomposition of the boundary $\partial\mathbf{\Omega}_s(d)$ as the connection of $\partial\mathbf{\Omega}_s^+(d)$ and $\partial\mathbf{\Omega}_s^-(d)$.

$$\{\angle(\boldsymbol{\Phi}\mathbf{u}, \boldsymbol{\Phi}\mathbf{v}) \mid \boldsymbol{\Phi}\mathbf{u} \text{ and } \boldsymbol{\Phi}\mathbf{v} \text{ satisfy } (2.2), (2.3), \text{ and } (2.7)\}$$

$$\stackrel{(a)}{=} \bigcup_{d_{\min} \leq d \leq d_{\max}} \{\angle B_d D_d C_d \mid D_d \in \boldsymbol{\Omega}(d)\} \stackrel{(b)}{=} \bigcup_{d_{\min} \leq d \leq d_{\max}} \{\angle BDC \mid D \in \boldsymbol{\Omega}_s(d)\} \stackrel{(c)}{=} \{\angle BDC \mid D \in \boldsymbol{\Omega}\}, \quad (2.25)$$

where (a) can be directly inferred from (2.19), (b) follows from (2.20), and (c) holds due to (2.24). The result of (2.25) then leads to the key theorem given below.

***Theorem 2.2:*** The following result holds:

$$\alpha_{\max} = \max_{D \in \boldsymbol{\Omega}} \angle BDC \quad \text{and} \quad \alpha_{\min} = \min_{D \in \boldsymbol{\Omega}} \angle BDC. \quad (2.26)$$

□

Theorem 2.2 provides the foundation behind the proposed approach. In particular, (2.26) asserts that we can focus on the single diagram as Figure 3, based on which plane geometry analyses can be then conducted to identify $\alpha_{\max}$ and $\alpha_{\min}$.

*C. Further Characterization of the Joint Feasible Top-Vertex Set $\boldsymbol{\Omega}$*

According to Theorem 2.2, all we have to do is to seek among all $D \in \boldsymbol{\Omega}$ for the two that will, respectively, yield the maximal and minimal $\angle BDC$ (cf. (2.26)). Intuitively speaking, $\alpha_{\max}$ and $\alpha_{\min}$ are very likely to be attained by some points located on $\partial \boldsymbol{\Omega}$, the boundary of $\boldsymbol{\Omega}$. By conducting plane geometry analyses based on the diagram in Figure 3, such a simple idea turns out to be true, as will be shown below. The result allows us to further narrow down the candidate top vertices in $\boldsymbol{\Omega}$ so as to simplify the identification of $\alpha_{\max}$ and $\alpha_{\min}$.

To proceed, we shall first provide more concrete characterization of $\partial \boldsymbol{\Omega}$; in particular, the left, right, top, and bottom components of $\partial \boldsymbol{\Omega}$ will be specified. For this, associated with each similar feasible top-vertex set $\boldsymbol{\Omega}_s(d)$ let

$$\eta_s(d) \triangleq \arg\min_{D \in \boldsymbol{\Omega}_s(d)} |\overline{BD}| \cos(\angle DBC) \quad \text{and} \quad \mu_s(d) \triangleq \arg\min_{D \in \boldsymbol{\Omega}_s(d)} |\overline{CD}| \cos(\angle DCB); \quad (2.27)$$

$\eta_s(d)$ and $\mu_s(d)$ represent, respectively, the left and right corner points of $\boldsymbol{\Omega}_s(d)$ (see Figure 4). Let us then decompose the boundary of $\boldsymbol{\Omega}_s(d)$, say $\partial \boldsymbol{\Omega}_s(d)$, into the connection of two branches as

$$\partial \boldsymbol{\Omega}_s(d) = \partial \boldsymbol{\Omega}_s^+(d) \cup \partial \boldsymbol{\Omega}_s^-(d), \quad (2.28)$$

where $\partial \boldsymbol{\Omega}_s^+(d)$ and $\partial \boldsymbol{\Omega}_s^-(d)$ are, respectively, the upper and lower branches both with $\eta_s(d)$ and $\mu_s(d)$ as the end points (see Figure 4). Since $\boldsymbol{\Omega}$ is obtained by overlaying all the continuum $\boldsymbol{\Omega}_s(d)$'s, with $\boldsymbol{\Omega}_s(d_{\min})$ on the top and $\boldsymbol{\Omega}_s(d_{\max})$ at the bottom (recall the construction of $\boldsymbol{\Omega}$), the upper branch of $\boldsymbol{\Omega}_s(d_{\min})$ and the lower branch of $\boldsymbol{\Omega}_s(d_{\max})$ must belong to $\partial \boldsymbol{\Omega}$, i.e., $\partial \boldsymbol{\Omega}_s^+(d_{\min}) \subset \partial \boldsymbol{\Omega}$ and $\partial\Omega_s^-(d_{\max}) \subset \partial\Omega$. This naturally suggests that the top and bottom boundaries of $\Omega$, denoted by $\partial\Omega_t$ and $\partial\Omega_b$, can be specified respectively as

$$\partial\Omega_t = \partial\Omega_s^+(d_{\min}) \tag{2.29}$$

and

$$\partial\Omega_b = \partial\Omega_s^-(d_{\max}). \tag{2.30}$$

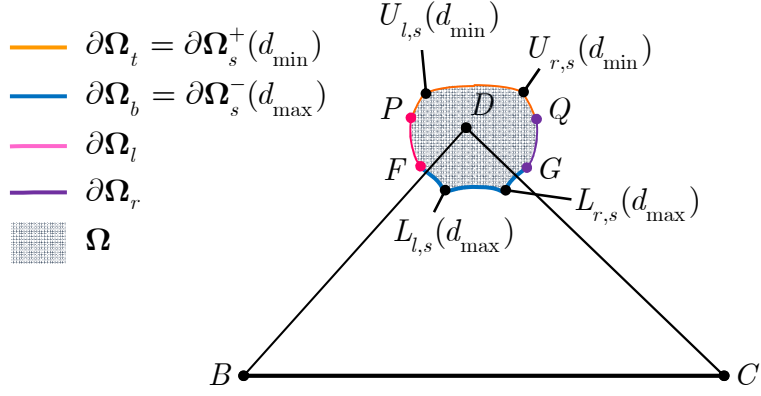

Figure 5. Depiction of the boundary of $\Omega$, in which $\partial\Omega_t$, $\partial\Omega_b$, $\partial\Omega_l$, and $\partial\Omega_r$ are, respectively, the top, bottom, left and right boundaries.

With $\partial\Omega_t$ and $\partial\Omega_b$ given as above, we can go on to determine the left and right boundaries of $\Omega$, denoted by $\partial\Omega_l$ and $\partial\Omega_r$ respectively. Indeed, $\partial\Omega_l$ is simply the boundary curve with end points $\eta_s(d_{\min})$ and $\eta_s(d_{\max})$, namely, the left ends of $\partial\Omega_t$ and $\partial\Omega_b$; similarly, $\partial\Omega_r$ is the portion with $\mu_s(d_{\min})$ and $\mu_s(d_{\max})$ as the two ends. Note that by construction of $\Omega$, it follows

$$\partial\Omega_l = \{\eta_s(d) \mid d_{\min} \leq d \leq d_{\max}\}, \text{ and } \partial\Omega_r = \{\mu_s(d) \mid d_{\min} \leq d \leq d_{\max}\}, \tag{2.31}$$

which are the collections of, respectively, the left and right corner points of $\Omega_s(d)$ for all plausible $d$. Hence, we have $\partial\Omega = \partial\Omega_r \cup \partial\Omega_l \cup \partial\Omega_t \cup \partial\Omega_b$; see Figure 5 for the depiction of the four boundary components, and in the sequel we use the shorthand $(P, Q, F, G)$ to denote the four end points $(\eta_s(d_{\min}), \mu_s(d_{\min}), \eta_s(d_{\max}), \mu_s(d_{\max}))$ in order to conserve notation. With the above characterizations of $\partial\Omega$, the next lemma asserts that all elements on $\partial\Omega_l$ and $\partial\Omega_r$, except the ends $(P, Q, F, G)$, can be excluded from the candidate set in regard to the identification of $\alpha_{\max}$ and $\alpha_{\min}$.

*Lemma 2.3:* The following results hold.
(1) $\angle BPC < \angle BDC < \angle BFC$ for all $D \in \partial\Omega_l \setminus \{P, F\}$.
(2) $\angle BQC < \angle BDC < \angle BGC$ for all $D \in \partial\Omega_r \setminus \{Q, G\}$.
*[Proof]:* See Appendix. □

In addition, it can be shown that $\alpha_{\max}$ and $\alpha_{\min}$ will never be attained by the points in the interior of $\Omega$. Hence, the candidate region consists of only the top and bottom boundaries of $\Omega$. More specifically, we have the following key theorem.

**Theorem 2.4:** $\alpha_{\max} = \angle BDC$ for some $D \in \partial\Omega_b$, and $\alpha_{\min} = \angle BD'C$ for some $D' \in \partial\Omega_t$.

*[Proof]:* See Appendix. □

Theorem 2.4 asserts that the candidate top vertices relevant to the identification of $\alpha_{\max}$ and $\alpha_{\min}$ can be narrowed down to those on, respectively, $\partial\Omega_b$ and $\partial\Omega_t$. Based on the established results, the derivations of $\alpha_{\max}$ and $\alpha_{\min}$ are given in the next two sections.

## III. DERIVATION OF $\alpha_{\max}$

In this section we will conduct the plane geometry analyses based on Figure 5 to derive $\alpha_{\max}$. According to Theorem 2.4 and (2.30), our task is to find the largest $\angle BDC$ among all $D \in \partial\Omega_s^-(d_{\max})$, which is the lower branch of $\Omega_s(d_{\max})$. The main results of this section are established in Theorems 3.2 and 3.4.

To proceed, we shall first highlight the main idea behind the proof. Let us decompose $\partial\Omega_s^-(d_{\max})$ as the union of three boundary curves $\mathcal{CV}(F, L_{l,s}(d_{\max}))$, $\mathcal{CV}(L_{l,s}(d_{\max}), L_{r,s}(d_{\max}))$ and $\mathcal{CV}(L_{l,s}(d_{\max}), G)$ as shown in Figure 5. We propose to first find the maximal $\angle BDC$ for $D \in \mathcal{CV}(L_{l,s}(d_{\max}), L_{r,s}(d_{\max}))$. The main advantages of such an approach are twofold. First, the cosine of $\angle BDC$ for $D \in \mathcal{CV}(L_{l,s}(d_{\max}), L_{r,s}(d_{\max}))$, as will be shown later, admits a very simple expression in terms of $|\overline{DB}|$ and $|\overline{DC}|$. With the aid of this expression, finding the maximal $\angle BDC$ for $D \in \mathcal{CV}(L_{l,s}(d_{\max}), L_{r,s}(d_{\max}))$ can be mathematically formulated as a constrained optimization problem which is analytically solvable. Second, even though the feasible set of such an optimization problem consists of only those $D \in \mathcal{CV}(L_{l,s}(d_{\max}), L_{r,s}(d_{\max}))$, the solution to this problem can act as a yardstick point, to which the achievable $\angle BDC$ for all $D \in \mathcal{CV}(F, L_{l,s}(d_{\max})) \cup \mathcal{CV}(L_{l,s}(d_{\max}), L_{r,s}(d_{\max}))$ $\cup \mathcal{CV}(L_{r,s}(d_{\max}), G) = \partial\Omega_s^-(d_{\max})$ can be readily compared via plane geometry analyses: this thus provides a unified and systematic way of identifying $\alpha_{\max}$.

Since by construction $\mathcal{CV}(L_{l,s}(d_{\max}), L_{r,s}(d_{\max}))$ is the bottom edge of $\Omega_s(d_{\max})$ (see Figure 2-(b) with $d = d_{\max}$), for any $D \in \mathcal{CV}(L_{l,s}(d_{\max}), L_{r,s}(d_{\max}))$ the sum of the squared length $|\overline{DB}|^2 + |\overline{DC}|^2$ must satisfy (cf. (2.22))

$$|\overline{DB}|^2 + |\overline{DC}|^2 = \left\{\frac{(\tilde{d}_{\min}^2(d_{\max}/d)^2 + d_{\max}^2)}{2}\right\}_{d=d_{\max}} = \frac{\tilde{d}_{\min}^2 + d_{\max}^2}{2}. \quad (3.1)$$

With (3.1) and since $|\overline{BC}| = d_{\max}$, we have

$$\cos(\angle BDC) = \frac{|\overline{DB}|^2 + |\overline{DC}|^2 - |\overline{BC}|^2}{2|\overline{DB}| \times |\overline{DC}|} = \frac{\frac{\tilde{d}_{\min}^2 + d_{\max}^2}{2} - d_{\max}^2}{2|\overline{DB}| \times |\overline{DC}|} = \frac{\tilde{d}_{\min}^2 - d_{\max}^2}{4|\overline{DB}| \times |\overline{DC}|}. \quad (3.2)$$

Hence, for $D \in \mathcal{CV}(L_{l,s}(d_{\max}), L_{r,s}(d_{\max}))$, $\cos(\angle BDC)$ depends on $\tilde{d}_{\min}^2 - d_{\max}^2$ as well as the product length $|\overline{DB}| \times |\overline{DC}|$. Since $\tilde{d}_{\min}^2 - d_{\max}^2$ is a constant independent of the top vertex $D$ (see (2.4) and (2.9)), to simplify the exposure we proceed to find the maximal $\angle BDC$, or the minimal $\cos(\angle BDC)$, by first considering the scenario $\tilde{d}_{\min}^2 - d_{\max}^2 \geq 0$.

A. *Case I:* $\tilde{d}_{\min}^2 - d_{\max}^2 \geq 0$

Since $\tilde{d}_{\min}^2 - d_{\max}^2 \geq 0$, equation (3.2) shows that minimization of $\cos(\angle BDC)$ amounts to maximizing the product $|\overline{DB}| \times |\overline{DC}|$. Observe that

$$\left(|\overline{DB}| + |\overline{DC}|\right)^2 = |\overline{DB}|^2 + |\overline{DC}|^2 + 2|\overline{DB}| \times |\overline{DC}| \stackrel{(a)}{=} \frac{\tilde{d}_{\min}^2 + d_{\max}^2}{2} + 2|\overline{DB}| \times |\overline{DC}|, \quad (3.3)$$

where (a) follows from (3.1). Equation (3.3) implies that, to maximize $|\overline{DB}| \times |\overline{DC}|$, it is equivalent to maximize the summed length $|\overline{DB}| + |\overline{DC}|$. As a result, minimization of $\cos(\angle BDC)$ among all $D \in \mathcal{CV}(L_{l,s}(d_{\max}), L_{r,s}(d_{\max}))$ under the assumption $\tilde{d}_{\min}^2 - d_{\max}^2 \geq 0$ can be formulated as a constrained optimization problem as

$$(P1)\ \textit{Maximize}\ |\overline{DB}| + |\overline{DC}|,\ \textit{subject to}\ \begin{cases} (i)\ |\overline{DB}|^2 + |\overline{DC}|^2 = (\tilde{d}_{\min}^2 + d_{\max}^2)/2, \\ (ii)\ \sqrt{1-\delta} \leq |\overline{DB}| \leq \sqrt{1+\delta}, \\ (iii)\ \sqrt{1-\delta} \leq |\overline{DC}| \leq \sqrt{1+\delta}. \end{cases}$$

Note that the three constraints in (P1) are indeed the characterization of $|\overline{DB}|$ and $|\overline{DC}|$ for $D \in \mathcal{CV}(L_{l,s}(d_{\max}), L_{r,s}(d_{\max}))$. The optimal $D$ obtained by solving the above optimization problem is given in the next lemma.

*Lemma 3.1:* Let $d_{\max}^2$ and $\tilde{d}_{\min}^2$ be defined in, respectively, (2.4) and (2.9). Assume that $\tilde{d}_{\min}^2 - d_{\max}^2 \geq 0$, and let $D_L \in \mathcal{CV}(L_{l,s}(d_{\max}), L_{r,s}(d_{\max}))$ be such that

$$|\overline{D_L B}| = |\overline{D_L C}| = \sqrt{(\tilde{d}_{\min}^2 + d_{\max}^2)/4}. \quad (3.4)$$

Then the optimal $(|\overline{DB}|, |\overline{DC}|)$ which solves (P1) is given by $(|\overline{D_L B}|, |\overline{D_L C}|)$. As a result, we have $\angle BD_L C > \angle BDC$ for all $D \in \mathcal{CV}(L_{l,s}(d_{\max}), L_{r,s}(d_{\max})) \setminus \{D_L\}$.

*[Proof]:* See Appendix. □

Roughly speaking, the closer the top vertex $D$ is to the bottom $\overline{BC}$, the larger the angle $\angle BDC$ will be. Since among the three boundary curves $\mathcal{CV}(F, L_{r,s}(d_{\max}))$, $\mathcal{CV}(L_{l,s}(d_{\max}), L_{r,s}(d_{\max}))$, and $\mathcal{CV}(L_{l,s}(d_{\max}), G)$, $\mathcal{CV}(L_{l,s}(d_{\max}), L_{r,s}(d_{\max}))$ is located nearer to $\overline{BC}$ (see Figure 5), it is expected that the maximally achievable $\angle BDC$ for $D \in \mathcal{CV}(L_{l,s}(d_{\max}), L_{r,s}(d_{\max}))$, namely, $\angle BD_L C$, is the global $\alpha_{\max}$. This turns out to be true, as shown in the following theorem.

**Theorem 3.2:** Let $d_{\max}^2$ and $\tilde{d}_{\min}^2$ be defined in, respectively, (2.4) and (2.9). Assume that $\tilde{d}_{\min}^2 - d_{\max}^2 \geq 0$. Then we have

$$\alpha_{\max} = \angle BD_L C = \cos^{-1}\left(\frac{\tilde{d}_{\min}^2 - d_{\max}^2}{\tilde{d}_{\min}^2 + d_{\max}^2}\right). \tag{3.5}$$

*[Proof]:* See Appendix. □

**B. Case II:** $\tilde{d}_{\min}^2 - d_{\max}^2 < 0$

Since $\tilde{d}_{\min}^2 - d_{\max}^2 < 0$, based on (3.2), (3.3), and by following similar arguments as in Section III-A, minimization of $\cos(\angle BDC)$ thus amounts to minimizing the summed length $|\overline{DB}| + |\overline{DC}|$. Specifically, the maximal $\angle BDC$ for all $D \in \mathcal{CV}(L_{l,s}(d_{\max}), L_{r,s}(d_{\max}))$ in this case can be obtained as the solution to the following constrained optimization problem

$$(P2)\ Minimize\ |\overline{DB}| + |\overline{DC}|,\ subject\ to\ \begin{cases}(i)\ |\overline{DB}|^2 + |\overline{DC}|^2 = (\tilde{d}_{\min}^2 + d_{\max}^2)/2,\\ (ii)\ \sqrt{1-\delta} \leq |\overline{DB}| \leq \sqrt{1+\delta},\\ (iii)\ \sqrt{1-\delta} \leq |\overline{DC}| \leq \sqrt{1+\delta}.\end{cases}$$

The optimal solution obtained by solving (P2) is given in the next lemma.

**Lemma 3.3:** Let $d_{\max}^2$ and $\tilde{d}_{\min}^2$ be defined in, respectively, (2.4) and (2.9). Assume that $\tilde{d}_{\min}^2 - d_{\max}^2 < 0$. Then the optimal $(|\overline{DB}|, |\overline{DC}|)$ which solves (P2) is given by $(|\overline{L_{l,s}(d_{\max})B}|, |\overline{L_{l,s}(d_{\max})C}|)$ or $(|\overline{L_{r,s}(d_{\max})B}|, |\overline{L_{r,s}(d_{\max})C}|)$. Hence, the maximal $\angle BDC$ among all $D \in \mathcal{CV}(L_{l,s}(d_{\max}), L_{r,s}(d_{\max}))$ is attained by $D = L_{l,s}(d_{\max})$ or $D = L_{r,s}(d_{\max})$.

*[Proof]:* See Appendix. □

With the aid of Lemma 3.3 and through further plane geometry analyses, the maximal angle in this case is derived in the next theorem.

**Theorem 3.4:** Let $d_{\max}^2$ and $\tilde{d}_{\min}^2$ be defined in, respectively, (2.4) and (2.9). Assume that $\tilde{d}_{\min}^2 - d_{\max}^2 < 0$. The following results hold.
(1) If $\tilde{d}_{\min}^2 + d_{\max}^2 \leq 4$, then

$$\alpha_{\max} = \cos^{-1}\left\{\max\left\{-1, \left(\frac{\tilde{d}_{\min}^2 - d_{\max}^2}{4\sqrt{1-\delta}\sqrt{(\tilde{d}_{\min}^2 + d_{\max}^2)/2 - (1-\delta)}}\right)\right\}\right\}. \tag{3.6}$$

(2) If $\tilde{d}_{\min}^2 + d_{\max}^2 > 4$, then

$$\alpha_{\max} = \cos^{-1}\left\{\max\left\{-1, \left(\frac{\tilde{d}_{\min}^2 - d_{\max}^2}{4\sqrt{(\tilde{d}_{\min}^2 + d_{\max}^2)/2 - (1+\delta)}\sqrt{1+\delta}}\right)\right\}\right\}. \tag{3.7}$$

*[Proof]:* See Appendix. □

## IV. DERIVATION OF $\alpha_{\min}$

By following the similar ideas and approaches as in Section III, we go on to derive $\alpha_{\min}$ in this section. According to Theorem 2.4 and (2.29), our task is to find the smallest $\angle BDC$ among all $D \in \partial\Omega_s^+(d_{\min})$, the top boundary of $\Omega$. The main results of this section are summarized in Theorems 4.2 and 4.4.

Let us likewise decompose $\partial\Omega_s^+(d_{\min})$ as the union of the three boundary curves $\mathcal{CV}(P, U_{l,s}(d_{\min}))$, $\mathcal{CV}(U_{l,s}(d_{\min}), U_{r,s}(d_{\min}))$ and $\mathcal{CV}(U_{r,s}(d_{\min}), Q)$ (see Figure 5). To identify $\alpha_{\min}$, we shall first find the smallest $\angle BDC$ for $D \in \mathcal{CV}(U_{l,s}(d_{\min}), U_{r,s}(d_{\min}))$, and further identify the minimal $\angle BDC$ for $D \in \partial\Omega_s^+(d_{\min}) = \mathcal{CV}(P, U_{l,s}(d_{\min})) \cup \mathcal{CV}(U_{l,s}(d_{\min}), U_{r,s}(d_{\min})) \cup \mathcal{CV}(U_{r,s}(d_{\min}), Q)$. Such an approach enjoys similar technical advantages as have been mentioned and evidenced in the previous section, wherein the closed-form formula of $\alpha_{\max}$ has been derived.

Note that $\mathcal{CV}(U_{l,s}(d_{\min}), U_{r,s}(d_{\min}))$ is the top edge of $\Omega_s(d_{\min})$ (see Figure 2-(b) with $d = d_{\min}$). Hence, for any $D \in \mathcal{CV}(U_{l,s}(d_{\min}), U_{r,s}(d_{\min}))$, the quantity $|\overline{DB}|^2 + |\overline{DC}|^2$ must satisfy (cf. (2.21))

$$|\overline{DB}|^2 + |\overline{DC}|^2 = \left(\frac{\tilde{d}_{\max}^2 + d^2}{2} \cdot \frac{d_{\max}^2}{d^2}\right)_{d=d_{\min}} = \frac{\tilde{d}_{\max}^2 + d_{\min}^2}{2} \cdot \xi^2, \tag{4.1}$$

where $\xi \triangleq d_{\max}/d_{\min}$ is used throughout this section to conserve notation. With (4.1) and since $|\overline{BC}| = d_{\max}$, we have

$$\cos(\angle BDC) = \frac{|\overline{DB}|^2 + |\overline{DC}|^2 - |\overline{BC}|^2}{2|\overline{DB}| \times |\overline{DC}|} = \frac{(\tilde{d}_{\max}^2 + d_{\min}^2)/2 - d_{\min}^2}{2|\overline{DB}| \times |\overline{DC}|} \cdot \xi^2 = \frac{\tilde{d}_{\max}^2 - d_{\min}^2}{4|\overline{DB}| \times |\overline{DC}|} \cdot \xi^2. \tag{4.2}$$

Hence, for $D \in \mathcal{CV}(U_{l,s}(d_{\min}), U_{r,s}(d_{\min}))$, $\cos(\angle BDC)$ depends on $(\tilde{d}_{\max}^2 - d_{\min}^2)\xi^2$ as well as the product length $|\overline{DB}| \times |\overline{DC}|$. Since $\xi^2(>0)$ and $\tilde{d}_{\max}^2 - d_{\min}^2$ are constants independent of $D$, to simplify the exposure we proceed to find the minimal $\angle BDC$, or the maximal $\cos(\angle BDC)$, by first considering the scenario $\tilde{d}_{\max}^2 - d_{\min}^2 < 0$.

## A. Case I: $\tilde{d}_{\max}^2 - d_{\min}^2 < 0$

As $\tilde{d}_{\max}^2 - d_{\min}^2 < 0$ and from (4.2), maximization of $\cos(\angle BDC)$ thus amounts to maximizing the product length $|\overline{DB}| \times |\overline{DC}|$. Since

$$\left(|\overline{DB}| + |\overline{DC}|\right)^2 = |\overline{DB}|^2 + |\overline{DC}|^2 + 2|\overline{DB}| \times |\overline{DC}| \stackrel{(a)}{=} \frac{\tilde{d}_{\max}^2 + d_{\min}^2}{2} \cdot \xi^2 + 2|\overline{DB}| \times |\overline{DC}|, \quad (4.3)$$

where (a) follows from (4.1), maximization of $|\overline{DB}| \times |\overline{DC}|$ is equivalent to maximize the summed length $|\overline{DB}| + |\overline{DC}|$. Hence, maximization of $\cos(\angle BDC)$ among all $D \in \mathcal{CV}(U_{l,s}(d_{\min}), U_{r,s}(d_{\min}))$ under the assumption $\tilde{d}_{\max}^2 - d_{\min}^2 < 0$ can be formulated as the following constrained optimization problem

$$(P3) \; Maximize \; |\overline{DB}| + |\overline{DC}|, \; \text{subject to} \; \begin{cases} (i) \; |\overline{DB}|^2 + |\overline{DC}|^2 = (\tilde{d}_{\max}^2 + d_{\min}^2) \cdot \xi^2 / 2, \\ (ii) \; \sqrt{1-\delta} \cdot \xi \leq |\overline{DB}| \leq \sqrt{1+\delta} \cdot \xi, \\ (iii) \; \sqrt{1-\delta} \cdot \xi \leq |\overline{DC}| \leq \sqrt{1+\delta} \cdot \xi. \end{cases}$$

Note that the three constraints in (P3) are the characterization of $|\overline{DB}|$ and $|\overline{DC}|$ for $D \in \mathcal{CV}(U_{l,s}(d_{\min}), U_{r,s}(d_{\min}))$. The optimal $D$ obtained by solving the above optimization problem is given in the next lemma.

***Lemma 4.1:*** Let $d_{\min}^2$ and $\tilde{d}_{\max}^2$ be defined in, respectively, (2.5) and (2.8). Assume that $\tilde{d}_{\max}^2 - d_{\min}^2 < 0$, and let $D_U \in \mathcal{CV}(U_{l,s}(d_{\min}), U_{r,s}(d_{\min}))$ be such that

$$|\overline{D_U B}| = |\overline{D_U C}| = \sqrt{(\tilde{d}_{\max}^2 + d_{\min}^2)/4} \cdot \xi. \quad (4.4)$$

Then the optimal $(|\overline{DB}|, |\overline{DC}|)$ which solves (P3) is given by $(|\overline{D_U B}|, |\overline{D_U C}|)$. Hence, we have $\angle BD_U C < \angle BDC$ for all $D \in \mathcal{CV}(U_{l,s}(d_{\min}), U_{r,s}(d_{\min})) \setminus \{D_U\}$.

*[Proof]:* See Appendix. □

Through further analyses it can be shown that $\angle BD_U C$ is the global minimal angle $\alpha_{\min}$, as established in the next theorem.

***Theorem 4.2:*** Let $d_{\min}^2$ and $\tilde{d}_{\max}^2$ be defined in, respectively, (2.5) and (2.8). Assume that $\tilde{d}_{\max}^2 - d_{\min}^2 < 0$. Then we have

$$\alpha_{\min} = \angle BD_U C = \cos^{-1}\left(\frac{\tilde{d}_{\max}^2 - d_{\min}^2}{\tilde{d}_{\max}^2 + d_{\min}^2}\right). \quad (4.5)$$

*[Proof]:* See Appendix. □

**B. Case II:** $\tilde{d}_{max}^2 - d_{min}^2 \geq 0$

Since $\tilde{d}_{max}^2 - d_{min}^2 \geq 0$ and from (4.2) and (4.3), similar arguments show that minimization of $\angle BDC$ for all $D \in \mathcal{CV}(U_{l,s}(d_{min}), U_{r,s}(d_{min}))$ in this case can be formulated as the following constrained optimization problem

$$(P4)\ Minimize\ |\overline{DB}| + |\overline{DC}|,\ subject\ to\ \begin{cases} (i)\ |\overline{DB}|^2 + |\overline{DC}|^2 = (\tilde{d}_{max}^2 + d_{min}^2) \cdot \xi^2 / 2, \\ (ii)\ \sqrt{1-\delta} \cdot \xi \leq |\overline{DB}| \leq \sqrt{1+\delta} \cdot \xi, \\ (iii)\ \sqrt{1-\delta} \cdot \xi \leq |\overline{DC}| \leq \sqrt{1+\delta} \cdot \xi. \end{cases}$$

The optimal $D$ obtained by solving (P4) is given in the next lemma.

***Lemma 4.3:*** Let $d_{min}^2$ and $\tilde{d}_{max}^2$ be defined in, respectively, (2.5) and (2.8). Assume that $\tilde{d}_{max}^2 - d_{min}^2 \geq 0$. Then the optimal $(|\overline{DB}|, |\overline{DC}|)$ which solves (P4) is given by $(|\overline{U_{l,s}(d_{min})B}|, |\overline{U_{l,s}(d_{min})C}|)$ or $(|\overline{U_{r,s}(d_{min})B}|, |\overline{U_{r,s}(d_{min})C}|)$. Hence, the minimal $\angle BDC$ among all $D \in \mathcal{CV}(U_{l,s}(d_{min}), U_{r,s}(d_{min}))$ is attained by $D = U_{l,s}(d_{min})$ or $D = U_{r,s}(d_{min})$.

*[Proof]:* See Appendix. □

With the aid of Lemma 4.3 and by means of plane geometry analyses, $\alpha_{min}$ in this case is derived in the following theorem.

***Theorem 4.4:*** Let $d_{min}^2$ and $\tilde{d}_{max}^2$ be defined in, respectively, (2.5) and (2.8). Assume that $\tilde{d}_{max}^2 - d_{min}^2 \geq 0$. The following results hold.

(1) Assume that $\tilde{d}_{max}^2 + d_{min}^2 < 4$. Then

$$\alpha_{min} = \cos^{-1}\left(\min\{1, \cos\alpha_1\}\right), \quad (4.6)$$

where

$$\cos\alpha_1 = \frac{\tilde{d}_{max}^2 - d_{min}^2}{4\sqrt{1-\delta}\sqrt{(\tilde{d}_{max}^2 + d_{min}^2)/2 - (1-\delta)}}. \quad (4.7)$$

(2) Assume that $\tilde{d}_{max}^2 + d_{min}^2 \geq 4$.

(a) If $(1+\delta) - d_{min}^2 \geq 0$, then

$$\alpha_{min} = \cos^{-1}\left(\min\{1, \cos\alpha_2\}\right), \quad (4.8)$$

where

$$\cos\alpha_2 = \begin{cases} \max\left[\dfrac{\tilde{d}_{\max}^2 - d_{\min}^2}{4\sqrt{1+\delta}\sqrt{(\tilde{d}_{\max}^2 + d_{\min}^2)/2 - (1+\delta)}}, \dfrac{\tilde{d}_{\min}^2 - d_{\min}^2}{4\sqrt{1+\delta}\sqrt{(\tilde{d}_{\min}^2 + d_{\min}^2)/2 - (1+\delta)}}\right] \\ \qquad\qquad\qquad\qquad\qquad\qquad\qquad\qquad\qquad\qquad\qquad\text{, if } \tilde{d}_{\min}^2 + d_{\min}^2 > 4; \\ \max\left\{\dfrac{\tilde{d}_{\max}^2 - d_{\min}^2}{4\sqrt{1+\delta}\sqrt{(\tilde{d}_{\max}^2 + d_{\min}^2)/2 - (1+\delta)}}, \dfrac{2 - d_{\min}^2}{2\sqrt{1+\delta}\sqrt{1-\delta}}\right\} \\ \qquad\qquad\qquad\qquad\qquad\qquad\qquad\qquad\qquad\qquad\qquad\text{, if } \tilde{d}_{\min}^2 + d_{\min}^2 \leq 4. \end{cases} \quad (4.9)$$

(b) If $(1+\delta) - d_{\min}^2 < 0$, then

$$\alpha_{\min} = \cos^{-1}\left(\min\{1, \cos\alpha_3\}\right) \quad (4.10)$$

where

$$\cos\alpha_3 = \frac{\tilde{d}_{\max}^2 - d_{\min}^2}{4\sqrt{1+\delta}\sqrt{(\tilde{d}_{\max}^2 + d_{\min}^2)/2 - (1+\delta)}} \quad (4.11)$$

*[Proof]:* See Appendix. □

## V. DISSCUSSIONS

*A. Connection to Previous Works*

It is known that a sharp upper bound for $|\cos\alpha|$ ($\alpha = \angle(\mathbf{\Phi u}, \mathbf{\Phi v})$) is crucial for accurate performance evaluations in many CS problems, e.g., [9], [19-20]. For the special case $\angle(\mathbf{u}, \mathbf{v}) = \theta = \pi/2$, thus $\cos\theta = 0$, a well-known upper bound for $|\cos\alpha|$ derived by means of the polarization identity is given by [9]

$$|\cos\alpha| \leq \min\left\{\frac{\delta}{1-\delta}, 1\right\}. \quad (5.1)$$

Using the similar algebraic approach as in [9], a generalization of the above result to the case when $\theta$ is arbitrary (but fixed) is given in the following theorem.

***Theorem 5.1:*** Assume that $\angle(\mathbf{u}, \mathbf{v}) = \theta$. The following inequality holds:

$$|\cos\alpha| = \frac{|\langle\mathbf{\Phi u}, \mathbf{\Phi v}\rangle|}{\|\mathbf{\Phi u}\|_2 \|\mathbf{\Phi v}\|_2} \leq \min\left\{\frac{\delta + |\cos\theta|}{1-\delta}, 1\right\}. \quad (5.2)$$

*[Proof]:* See Appendix. □

Now, with the aid of the proposed $\alpha_{\max}$ and $\alpha_{\min}$ in the previous subsections, the achievable $|\cos\alpha|$ can be directly determined as

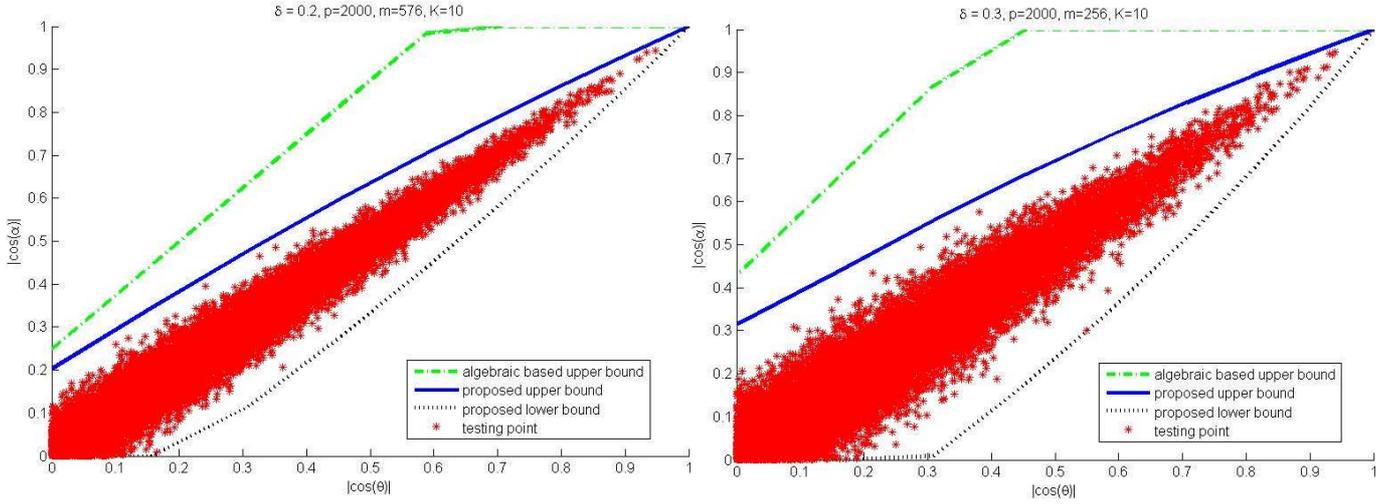

Figure 6. Achievable $|\angle\cos(\alpha)|$ obtained by the proposed solution (5.3) and the upper bound (5.2) derived using the polarization identity.

$$\begin{cases} |\cos(\alpha_{\max})| \leq |\cos(\alpha)| \leq |\cos(\alpha_{\min})|, & \text{if } 0 < \alpha_{\min} < \alpha_{\max} \leq \pi/2, \\ |\cos(\alpha_{\min})| \leq |\cos(\alpha)| \leq |\cos(\alpha_{\max})|, & \text{if } \pi/2 < \alpha_{\min} < \alpha_{\max} < \pi, \\ 0 \leq |\cos(\alpha)| \leq \max\{|\cos(\alpha_{\min})|, |\cos(\alpha_{\max})|\}, & \text{if } 0 < \alpha_{\min} < \pi/2 < \alpha_{\max} < \pi. \end{cases} \quad (5.3)$$

For $\delta = 0.2, \ 0.3$, Figure 6 compares the upper bound in (5.2) and the one given by (5.3) for different $\theta$. It can be seen that our solution (5.3) is tighter than (5.2). This is not unexpected, since the bound (5.2) obtained via the algebraic-based approach is the worst-case estimate. The proposed bounds in (5.3) are further corroborated by computer simulations. To generate the test samples, the entries of the sensing matrix $\mathbf{\Phi} \in \mathbb{R}^{m \times p}$ are independently drawn from $\mathcal{N}(0, 1/m)$, namely, the Gaussian distribution with zero mean and variance equal to $1/m$. The ambient signal dimension is set to be $p = 2000$, and the sparsity level is $K = 10$. To guarantee $\mathbf{\Phi}$ satisfies RIP with an RIC equal to $\delta$ (with a high probability), the required measurement size is set in accordance with $m \sim O(K \log(p/K)/\delta^2)$ [3]. Associated with each $\delta$, a total number of 20000 $K$-sparse vector pairs $\{(\mathbf{u}_i, \mathbf{v}_i) \,|\, 1 \leq i \leq 20000\}$ are generated. For each test pair $(\mathbf{u}_i, \mathbf{v}_i)$, the angles $\theta_i = \angle(\mathbf{u}_i, \mathbf{v}_i)$ and $\alpha_i = \angle(\mathbf{\Phi}\mathbf{u}_i, \mathbf{\Phi}\mathbf{v}_i)$ are computed and then plotted on Figure 6. As we can see, the proposed solutions (5.3) are indeed tight estimates of the achievable $\alpha_i$'s.

*B. Special Case* $\angle(\mathbf{u}, \mathbf{v}) = \theta = \pi/2$

For the special case $\angle(\mathbf{u}, \mathbf{v}) = \theta = \pi/2$, the closed-form formulae of $\alpha_{\max}$ and $\alpha_{\min}$ provided in Sections III and IV can be considerably simplified. Specifically, when $\theta = \pi/2$, it is easy to verify that

$$d_{\max}^2 = \tilde{d}_{\max}^2 = 2(1+\delta) \text{ and } d_{\min}^2 = \tilde{d}_{\min}^2 = 2(1-\delta). \quad (5.4)$$

Based on (5.4) together with some straightforward manipulations, we have the following corollary, which will be used to obtain improved performance guarantees in several CS systems in the next section.

*Corollary 5.2:* Assuming that $\angle(\mathbf{u},\mathbf{v}) = \pi/2$, $\alpha_{\max}$ and $\alpha_{\min}$ are, respectively, given as

$$\alpha_{\max} = \cos^{-1}\left(\max\left\{-1, \frac{-\delta}{\sqrt{1-\delta^2}}\right\}\right), \tag{5.5}$$

and

$$\alpha_{\min} = \cos^{-1}\left(\min\left\{1, \frac{\delta}{\sqrt{1-\delta^2}}\right\}\right). \tag{5.6}$$

□

# VI. APPLICATIONS

Consider the following CS system, in which the effective sensing matrix is the product of an orthogonal projection matrix $\mathbf{P}$ and the original sensing matrix $\mathbf{\Phi}$:

$$\mathbf{y} = \mathbf{P}\mathbf{\Phi}\mathbf{x}, \tag{6.1}$$

where $\mathbf{P} \triangleq \mathbf{I} - \mathbf{\Phi}_{T_I}(\mathbf{\Phi}_{T_I}^*\mathbf{\Phi}_{T_I})^{-1}\mathbf{\Phi}_{T_I}^*$, with $\mathbf{I}$ being the identity matrix and $\mathbf{\Phi}_{T_I}$ consisting of the columns of $\mathbf{\Phi}$ indexed by a known set $T_I$. Note that the considered $\mathbf{P}$ acts as the orthogonal projection onto the orthogonal complement of the column space of $\mathbf{\Phi}_{T_I}$. The system model (6.1) arises, e.g., in compressed-domain interference cancellation [9-10], in modeling the residual vector of the OMP algorithm [19], and in establishing the democratic property of random sensing matrices [21-22]. Characterization of the RIC of $\mathbf{P}\mathbf{\Phi}$ is crucial for performance evaluation [9-10, 21-22] and for the study of sufficient signal reconstruction conditions [19]. Based on the derived $\alpha_{\max}$ and $\alpha_{\min}$ in Sections III and IV, the RIC of $\mathbf{P}\mathbf{\Phi}$ is specified in the following theorem.

*Theorem 6.1:* Consider the CS system (6.1). Assume that $\mathbf{\Phi}$ satisfies the RIP of order $K$ with RIC given by $\delta$. Let $T_I$ be an index set with $|T_I| < K$, i.e., the cardinality of $T_I$ is less than $K$. The following inequality holds for all $(K-|T_I|)$-sparse $\mathbf{x}$ whose support does not overlap with $T_I$:

$$(1-\bar{\delta})\|\mathbf{x}\|_2^2 \leq \|\mathbf{P}\mathbf{\Phi}\mathbf{x}\|_2^2 \leq (1+\delta)\|\mathbf{x}\|_2^2, \tag{6.2}$$

where

$$\bar{\delta} = \min\left\{1, \delta + \frac{\delta^2}{1+\delta}\right\}. \tag{6.3}$$

*[Proof]:* See Appendix. □

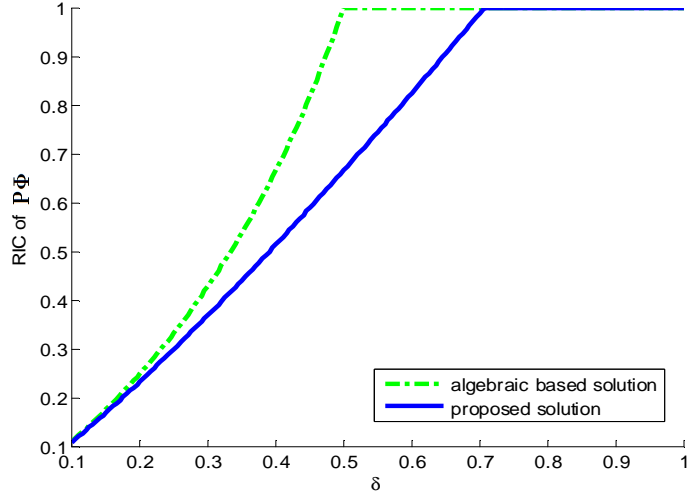

Figure 7. Comparison of the RIC of the effective sensing matrix $\mathbf{P}\boldsymbol{\Phi}$.

Theorem 6.1 asserts that $\mathbf{P}\boldsymbol{\Phi}$ satisfies RIP with an RIC equal to $\bar{\delta}$ given in (6.3). Under the same assumptions as in Theorem 6.1 and by means of the polarization identity, an inequality analogous to (6.2) has been derived in [9] and [19] as:

$$\left(1 - \frac{\delta}{1-\delta}\right)\|\mathbf{x}\|_2^2 \leq \|\mathbf{P}\boldsymbol{\Phi}\mathbf{x}\|_2^2 \leq (1+\delta)\|\mathbf{x}\|_2^2, \tag{6.4}$$

where $\mathbf{x}$ is $(K-|T_I|)$-sparse whose support does not overlap with $T_I$. Inequality (6.4) asserts that the RIC of $\mathbf{P}\boldsymbol{\Phi}$ is equal to

$$\bar{\delta}_a = \min\left\{1, \frac{\delta}{1-\delta}\right\}. \tag{6.5}$$

Since

$$\frac{\delta}{1-\delta} = \delta + \frac{\delta^2}{1-\delta}, \tag{6.6}$$

it is easy to see from (6.3) and (6.6) that $\bar{\delta} < \bar{\delta}_a$, that is, the proposed RIC $\bar{\delta}$ in (6.3) is tighter. The numerical values of the proposed solution (6.3) and the algebraic-based estimate $\bar{\delta}_a$ in (6.5) with respect to different $\delta$ are computed and plotted in Figure 7. It is seen that considerable improvement can be achieved by our solution for moderate[6] $\delta$. Applications of the above results to three CS problems are discussed below.

*A. Compressed-Domain Interference Cancellation*

In this problem, equation (6.1) represents the effective data acquisition system after the undesirable interference is removed via orthogonal projection [9-10]. A tighter RIC of the effective sensing matrix $\mathbf{P}\boldsymbol{\Phi}$ can make impacts on the following two aspects.

*i) Improved Signal Reconstruction Performance Evaluation:* It is well-known that a small RIC of the sensing matrix $\mathbf{P}\boldsymbol{\Phi}$ yields a better signal reconstruction performance [1-3]. Specifically, assume that the commonly used $l_1$-minimization algorithm [17-18] is adopted for signal recovery, and let $\hat{\mathbf{x}}$ be the reconstructed sparse signal vector. Based on [22, eq. (5) and (6)], the upper bounds of the reconstruction error determined in accordance with the proposed RIC $\bar{\delta}$ in (6.3) and the algebraic-based estimate $\bar{\delta}_a$ in (6.5) are, respectively[7],

$$\|\hat{\mathbf{x}} - \mathbf{x}\|_2 \leq \frac{4(1+\bar{\delta})\varepsilon}{1-(\sqrt{2}-1)\bar{\delta}} = \frac{4\left[1+\delta+\delta/(1+\delta^2)\right]\varepsilon}{1-(\sqrt{2}-1)\left[\delta+\delta/(1+\delta^2)\right]}, \tag{6.7}$$

and

$$\|\hat{\mathbf{x}} - \mathbf{x}\|_2 \leq \frac{4(1+\bar{\delta}_a)\varepsilon}{1-(\sqrt{2}-1)\bar{\delta}_a} = \frac{4(1+\delta/(1-\delta))\varepsilon}{1-(\sqrt{2}-1)(\delta/(1-\delta))}, \tag{6.8}$$

where $\varepsilon$ is the level of the data mismatch (measured in $l_2$-norm). For $\varepsilon = 0.01$, Figure 8 plots the error upper bounds in (6.7) and (6.8) as a function of $\delta$. It can be seen that the proposed solution does yield a tighter estimate of the reconstruction error; for moderate $\delta$ (say, $\delta > 0.2$) the improvement can be as large as $4.7$ dB.

---

6. A large $\delta$ will result in the failure of signal recovery, and thus should be precluded [24].

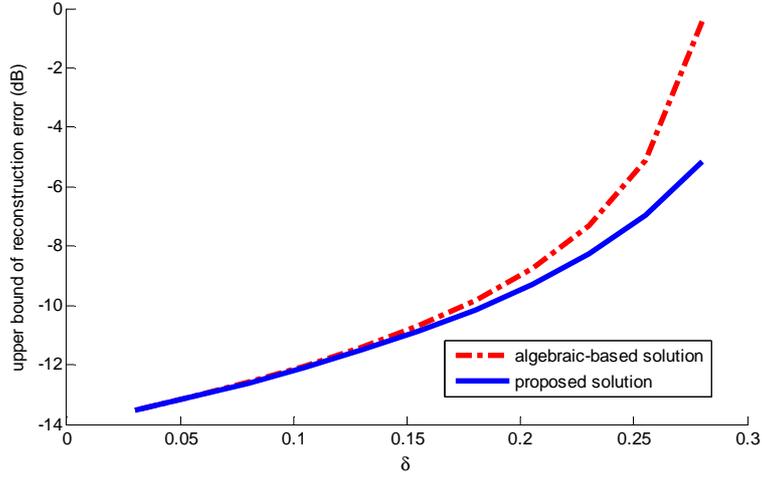

Figure 8. Upper bound of signal reconstruction error with respect to different $\delta$.

*(ii) Reduction in the Measurement Size:* Given a target signal reconstruction performance upon interference removal, a tighter RIC of $\mathbf{P\Phi}$ is also potentially conducive to the reduction in the measurement size. To be specific, assume that a threshold $\tau$ of the RIC of $\mathbf{P\Phi}$ is imposed for a certain performance guarantee. The required $\delta$ (the RIC of $\mathbf{\Phi}$) determined according to the proposed solution (6.3) and the algebraic based estimate (6.5) are, respectively, allowed to be as large as

$$\delta = \frac{\tau - 1 + \sqrt{\tau^2 + 6\tau + 1}}{4}, \tag{6.9}$$

and

---

7. For (6.7) and (6.8) to hold, it is assumed that the RIC of $\mathbf{P\Phi}$ does not exceed $\sqrt{2}-1$, the sufficient condition for stable signal recovery (see [17], [22]).

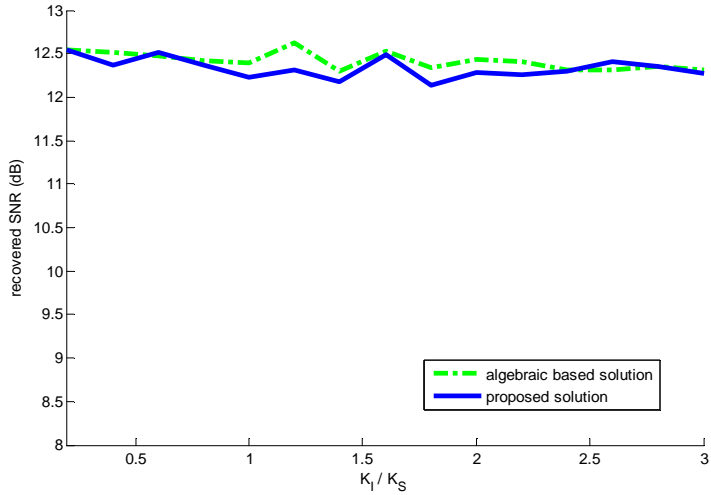

Figure 9. Recovered SNR based on two solutions.

$$\delta = \frac{\tau}{\tau+1}. \qquad (6.10)$$

It can be verified that the one computed based on our solution (i.e., (6.9)) is larger. It is well-known that, with random sensing and under a fixed ambient signal dimension $p$, the required number of measurements is inversely proportional to $\delta^2$ [3]. Hence, our solution guarantees that successful signal recovery can be achieved with fewer measurements. As an example, let us pick $\tau = 0.4$ ($<\sqrt{2}-1=0.414$) so that the sufficient condition for stable signal reconstruction using the $l_1$-minimization algorithm [17-18] holds. By computations, it is found that the proposed solution (6.9) yields $\delta = 0.3208$, whereas the algebraic based solution (6.10) leads to a pessimistic estimate as small as $\delta = 0.2843$. Let the entries of $\Phi$ be drawn from $\mathcal{N}(0,1/m)$, where the measurement size $m \sim O(K\log(p/K)/\delta^2)$; then, with an overwhelming probability, $\Phi$ fulfills the RIP of order $K$ with the prescribed RIC $\delta$ [3]. As such, the proposed solution can achieve about a $21.5\%$ reduction in the number of measurements.

To further demonstrate the resultant signal recovery performance, we consider two sensing systems both with the ambient signal dimension $p = 1000$ and $K = 20$; the numbers of measurements are set to be, respectively, $m = 661$ and $m = 841$ so that the RIC of the sensing matrix $\Phi$ can achieve $\delta = 0.3208$ and $\delta = 0.2843$ with a high probability. We use the algorithm proposed in [17-18] for signal reconstruction; since the level of the $l_2$-norm data error must be known in the algorithm formulation (see [17-18]), the sensing data mismatch is modeled as a uniform random variable over $[-\sigma, \sigma]$, where $\sigma$ is determined according to the SNR in the received data. The cardinality of the desired sparse signal vector is fixed to be $K_S = 5$, whereas the cardinality of interference $K_I$ fulfills $K_S + K_I \leq 20$. Figure 9 plots the recovered SNR [21] as a function of $(K_I/K_S)$ (the background SNR is 15 dB). The figure shows that the recovered SNR achieved by the two sensing systems are pretty close; as one can see, a $21.5\%$ reduction in the number of measurements merely incurs an SNR degradation no more than $0.5$ dB.

*B. RIP-Based Analysis of OMP Algorithm*

The signal model (6.1) also arises in the modeling of the residual vector of the OMP algorithm so as to facilitate stability analysis [19]. Based on (6.4), it was shown in [19] that the OMP algorithm can successfully recover any $K$-sparse vector provided that the sensing matrix $\Phi$ satisfies RIP of order $K+1$ with an RIC $\delta < 1/(3\sqrt{K})$. In [20], an improved sufficient condition was obtained, and is stated in the next proposition.

***Proposition 6.2 ([20]):*** Assume that the sensing matrix $\Phi$ satisfies RIP of order $K+1$ with RIC

$$\delta < 1/(1+\sqrt{2K}). \tag{6.11}$$

Then the OMP algorithm can recover a $K$-sparse vector in exactly $K$ iterations. □

Based on the tighter RIC of $\mathbf{P}\Phi$ derived in (6.3), an alternative sufficient condition is given in the next theorem.

***Theorem 6.3:*** Assume that the sensing matrix $\Phi$ satisfies RIP of order $K+1$ with RIC

$$\delta < \frac{1 - 2\sqrt{K} + \sqrt{4K + 12\sqrt{K} + 1}}{8\sqrt{K}}. \tag{6.12}$$

The OMP algorithm can recover a $K$-sparse vector in exactly $K$ iterations.

*[Proof]:* See Appendix. □

Figure 10 (see the next page) compares the required $\delta$ characterized by the upper bounds in (6.11) and (6.12) for different sparsity level $K$. It is seen that the required $\delta$ asserted by the proposed solution (6.12) is larger than that given by (6.11). Hence, in contrast with [20], Theorem 6.3 provides a less restricted sufficient condition for exact signal recovery when using OMP algorithm. We go on to illustrate, and compare, the signal reconstruction performances of two sensing systems with RIC specified by (6.11) and (6.12), respectively. The ambient signal size is set to be $p = 1000$, and $\Phi$ is a Gaussian random sensing matrix. For sparsity level $1 \leq K \leq 10$, the required $\delta$ are, respectively, computed using (6.11) and (6.12); the measurement size is then determined according to $m \sim O(K \log(p/K)/\delta^2)$ [3]. Figure 11-(a) compares the resultant probabilities of exact recovery for different $K$. It shows that the proposed sensing system, though with a larger $\delta$, remains capable of achieving perfect signal reconstruction. It is noted that a larger $\delta$ allows for a reduced measurement size [3]. For $1 \leq K \leq 10$, Figure 11-(b) further shows the resultant percentage of reduction in the number of measurements achieved by the proposed solution (6.12). As can be seen from the figure, the reduction is large when $K$ is small.

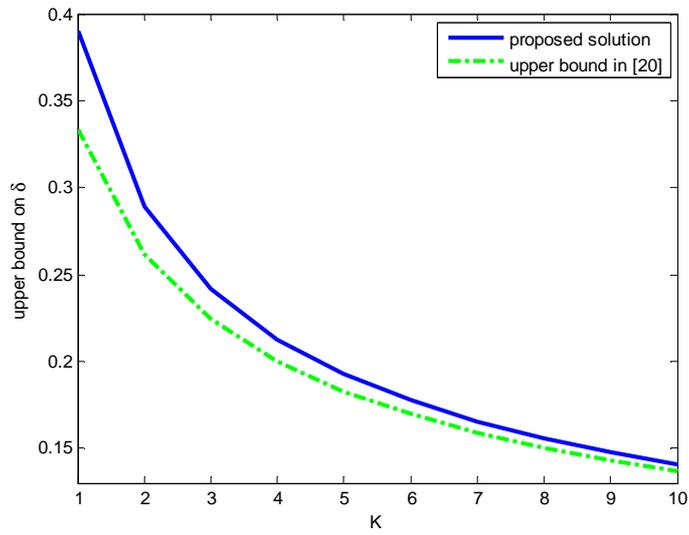

Figure 10. Required RIC $\delta$ obtained by (6.11) and (6.12) at various sparsity level $K$.

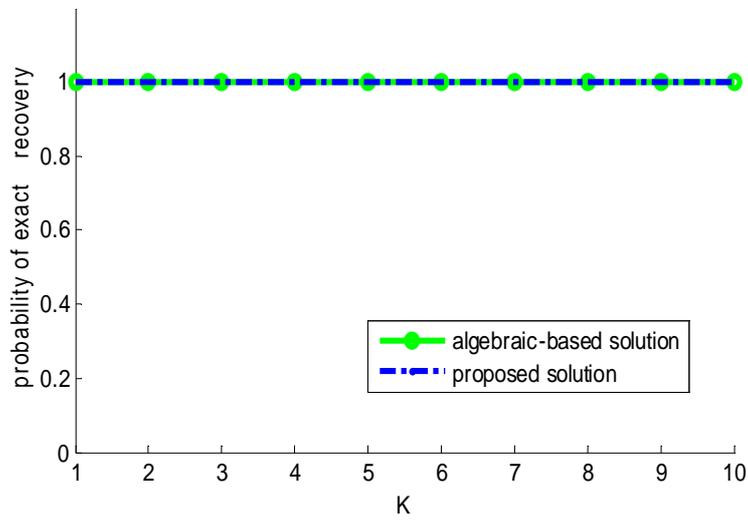

(a)

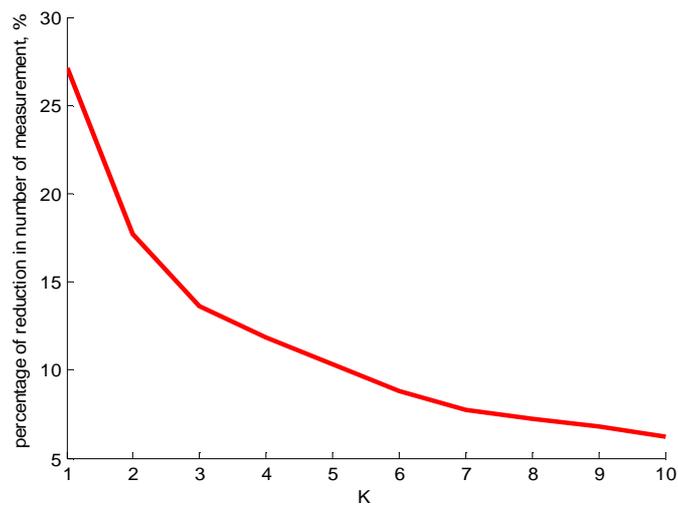

(b)

Figure 11. (a) Probability of exact recovery; (b) the percentage of reduction in the measurement size that is achieved by the proposed solution (6.12).

Notably, Figure 11-(a) together with Figure 11-(b) confirms that the conservation of measurements (as allowed by the less restricted sufficient condition shown in Theorem 6.3) does not incur any loss in the signal recovery performance.

*C. Democracy of Random Sensing Matrices*

The RIC of $\mathbf{P\Phi}$ is also needed in establishing the democratic property of random sensing matrices [21-22]; such an appealing characteristic guarantees the robustness of random data acquisition against the loss of measurements [21]. Let $\tilde{\mathbf{\Phi}} \in \mathbb{R}^{M \times N}$, with entries drawn from $\mathcal{N}(0, 1/M)$, satisfy RIP of order $K$ with RIC given by $\delta$. By democracy it is meant that, as long as $M$ is large enough, a sub-matrix obtained by deleting a small, and randomly chosen, subset of rows of $\tilde{\mathbf{\Phi}}$ still satisfies RIP but with a larger RIC. More precisely, assume that (i) $\tilde{M}$ is a positive integer such that $\tilde{M} \leq M$, $K \leq \tilde{M}$, and $D = M - \tilde{M} \geq 0$, and (ii) $M = C_1(K+D)\log((N+M)/(K+D))$ for some constant $C_1$. Then, with a high probability, a sub-matrix $\tilde{\mathbf{\Phi}}_{\tilde{M}} \in \mathbb{R}^{\tilde{M} \times N}$ obtained by removing arbitrary $D$ rows of $\tilde{\mathbf{\Phi}}$ satisfies RIP of order $K$ with RIC equal to $\delta/(1-\delta)$ [21]. By following the same proof procedures as in [21], it can be directly verified that the RIC of $\tilde{\mathbf{\Phi}}_{\tilde{M}}$ can be further tightened to $\bar{\delta}$ given in (6.3) (recall from Figure 7 that $\bar{\delta}$ is uniformly smaller than $\delta/(1-\delta)$). Note that $\tilde{\mathbf{\Phi}}_{\tilde{M}}$ represents the sensing matrix of the effective sensing system when $D$ measurements are dropped [21]. The established result, namely, $\tilde{\mathbf{\Phi}}_{\tilde{M}}$ enjoys a smaller RIC, confirms that better robustness of random sensing against measurement loss can be achieved.

# VII. CONCLUSIONS

In this paper, the achievable angles between two compressed sparse vectors under RIP-induced norm/distance constraints are analytically characterized in a plane geometry framework. Motivated by the law of cosines and geometric interpretations of RIP, it is shown that all the algebraic constraints imposed by RIP that are pertinent to the identification of the achievable angles can be jointly depicted via a simple geometric diagram in the two-dimensional plane. The proposed formulation allows for a unified joint analysis of all the considered algebraic constraints from a geometric point of view. By conducting plane geometry analyses based on the constructed diagram, the maximal and minimal angles can be derived in closed-form, and are then corroborated by numerical simulations. Compared with the existing algebraic based method employing the polarization identity, the proposed approach does provide sharper estimates of the achievable angles. Applications of our study to CS are investigated. First of all, we derive a closed-form RIC of the product of an orthogonal projection matrix and a random sensing matrix. Our solution is shown to be tighter than an RIC estimate reported in the literature. An immediate application of the result above is in compressed-domain interference cancellation. Specifically, we can provide a tighter estimate of the signal reconstruction error; in addition, it is shown that the required measurement size for achieving a target signal reconstruction quality can be further reduced. Regarding stability analysis of the OMP algorithm, a less restricted sufficient condition, which guarantees the OMP algorithm to perfectly recover a $K$-sparse vector in $K$ iterations, is established. The condition asserts that a larger RIC of the original sensing matrix is allowed: this guarantees exact signal recovery with a reduced measurement size. As the last application, we show that, with a small randomly chosen subset of rows removed from a random sensing matrix, the resultant sub-matrix enjoys an RIC smaller than the one reported in the literature. The result asserts that random sensing can provide better robustness against the measurement loss. It is believed that the presented analytic study of the achievable angles between two compressed sparse vectors can be of fundamental importance in the study of many other CS problems; this is currently under investigation.

# APPENDIX: SUPPORTING TECHNICAL PROOFS

The following two lemmas will be used throughout the appendix.

***Lemma A.1***: Consider the two triangles $\triangle KJL$ and $\triangle KSL$ in Figure A.1. Then $\angle KJL < \angle KSL$.

*[Proof]:* The result follows since

$$\angle KSL = \pi - (\angle SKL + \angle SLK) = \pi - (\pi - \angle JKS - \angle JLS - \angle KJL)$$
$$= \angle JKS + \angle JLS + \angle KJL > \angle KJL.$$

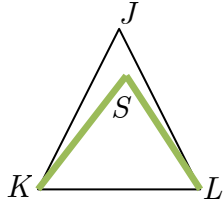

Figure A.1. Depiction of two triangles considered in Lemma A.1.

□

***Lemma A.2:*** Let $\overline{BC}$ be a chord of a circle. Pick $V$ and $W$ as two points above $\overline{BC}$, with $V$ inside the circle and $W$ outside the circle as depicted in Figure A.2. Also, let $A$ be an arbitrary point on the arc of the circle above $\overline{BC}$. Then $\angle BWC < \angle BAC < \angle BVC$.

*[Proof]:* The triangle $\triangle WBC$ intersects with the circle, say, at $B'$ and $C'$. Let $A'$ be a point on arc $B'C'$. By lemma A.1, we have

$$\angle BWC < \angle BA'C. \tag{A.1}$$

Let's further pick $A''$ as a point on the circle such that $\triangle BVC$ lies entirely inside $\triangle BA''C$. Again by Lemma A.1 we have

$$\angle BA''C < \angle BVC. \tag{A.2}$$

A well-known result from plane geometry is that angles inscribed in the same arc of a circle are equal [20], thus

$$\angle BA'C = \angle BAC = \angle BA''C. \tag{A.3}$$

The assertion follows from (A.1), (A.2), and (A.3). □

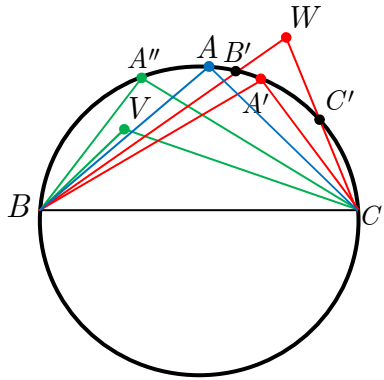

Figure A.2. Depiction of geometric objects considered in Lemma A.2.

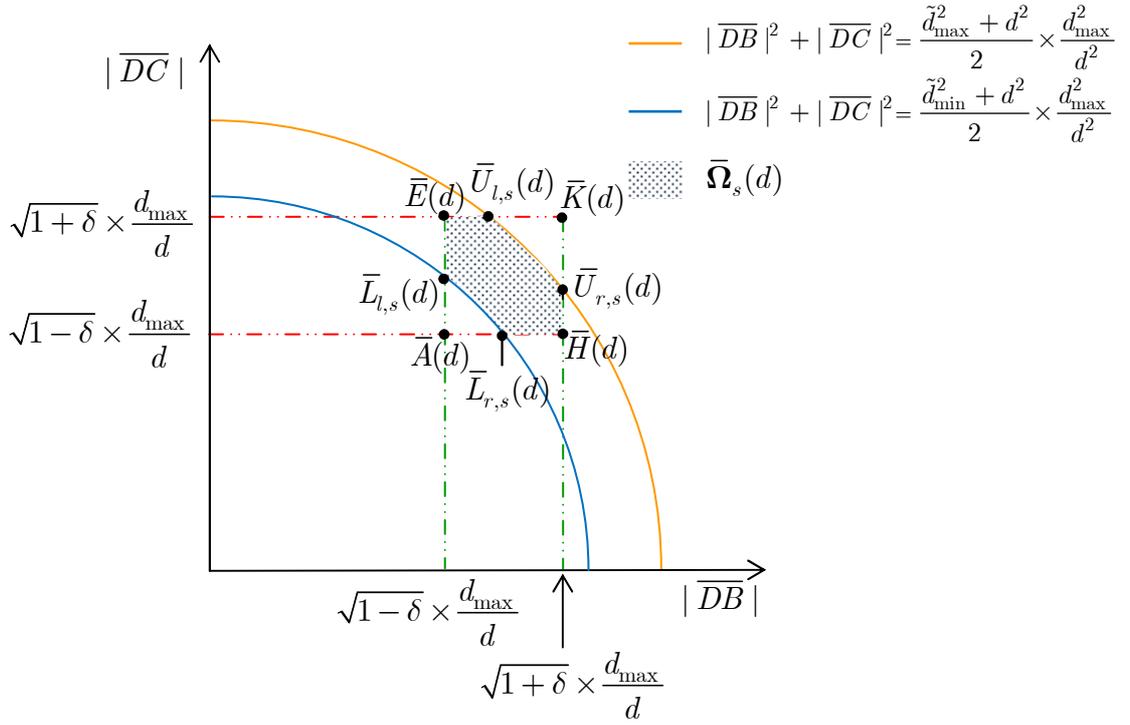

Figure A.3. An alternative depiction of the similar feasible top-vex set $\boldsymbol{\Omega}_s(d)$ via the considered coordinate system.

Throughout the appendix we will frequently leverage an alternative schematic description of the similar feasible top-vertex set $\boldsymbol{\Omega}_s(d)$ to ease analysis. To construct this diagram, for $D \in \boldsymbol{\Omega}_s(d)$ we shall first specify the constraints on the dilated length $|\overline{DB}|$ and $|\overline{DC}|$ under RIP. Pick a plausible $\|\boldsymbol{\Phi}(\mathbf{u}-\mathbf{v})\|_2 = d$. For each $D \in \boldsymbol{\Omega}_s(d)$ in Figure 2-(b), there is one and only one corresponding feasible top vertex $D_d \in \boldsymbol{\Omega}(d)$ in Figure 2-(a). Since Figure 2-(b) is obtained as the dilation of Figure 2-(a) with the scale $d_{\max}/d$, we have $|\overline{DB}| = |\overline{D_d B_d}| \times (d_{\max}/d)$ and $|\overline{DC}| = |\overline{D_d C_d}| \times (d_{\max}/d)$. Hence, for $D \in \boldsymbol{\Omega}_s(d)$ the norm constraints on $|\overline{DB}|$ and $|\overline{DC}|$ can be directly modified based on (2.2) as

$$\sqrt{1-\delta} \times \frac{d_{\max}}{d} \leq |\overline{DB}| \leq \sqrt{1+\delta} \times \frac{d_{\max}}{d}, \quad \sqrt{1-\delta} \times \frac{d_{\max}}{d} \leq |\overline{DC}| \leq \sqrt{1+\delta} \times \frac{d_{\max}}{d}; \qquad (A.4)$$

similarly, the constraint on $|\overline{DB}|^2 + |\overline{DC}|^2$ can be obtained from (2.13) as

$$\frac{\tilde{d}_{\min}^2 + d^2}{2} \times \frac{d_{\max}^2}{d^2} \leq |\overline{DB}|^2 + |\overline{DC}|^2 \leq \frac{\tilde{d}_{\max}^2 + d^2}{2} \times \frac{d_{\max}^2}{d^2}. \qquad (A.5)$$

Now, we consider the first quadrant of the coordinate system in the plane, with $|\overline{DB}|$ as the abscissa and $|\overline{DC}|$ as the ordinate. Taking into account (A.4) and (A.5), the depiction of all $D \in \boldsymbol{\Omega}_s(d)$ alternatively on this coordinate system is thus shown as the shadowed region $\bar{\boldsymbol{\Omega}}_s(d)$ in Figure A.3, in which $\bar{U}_{l,s}(d)$, $\bar{U}_{r,s}(d)$, $\bar{L}_{l,s}(d)$ and $\bar{L}_{r,s}(d)$ correspond to, respectively, the four points $U_{l,s}(d)$, $U_{r,s}(d)$, $L_{l,s}(d)$ and $L_{r,s}(d)$ on $\boldsymbol{\Omega}_s(d)$. The main advantage of the alternative depiction $\bar{\boldsymbol{\Omega}}_s(d)$ in

Figure A.3 is that it allows for a simple way of specifying $|\overline{DB}|$ and $|\overline{DC}|$ associated with all $D \in \mathbf{\Omega}_s(d)$. This will in turn simplify the underlying analyses, as will be seen later.

*A. Proof of Lemma 2.3*

Owing to the symmetric nature of $\mathbf{\Omega}$, it suffices to prove the first assertion. Since $\partial \mathbf{\Omega}_l$ consists of all the left end points of $\mathbf{\Omega}_s(d)$'s, $d_{\min} \leq d \leq d_{\max}$ (see (2.31)), it remains to show that, as $d$ increases from $d_{\min}$ to $d_{\max}$, $\angle B\eta_s(d)C$ is monotonically increasing, or equivalently, $\cos(\angle B\eta_s(d)C)$ is monotonically decreasing. Toward this end, we shall first find an explicit expression for $\cos(\angle B\eta_s(d)C)$ as a function of $d$. From the law of cosines (2.1), only $|\overline{\eta_s(d)B}|$ and $|\overline{\eta_s(d)C}|$ are needed to determine $\cos(\angle B\eta_s(d)C)$ (recall that $|\overline{BC}|$ is set to be $|\overline{BC}|= d_{\max}$). To ease the derivation, we will resort to the diagram $\overline{\mathbf{\Omega}}_s(d)$ in Figure A.3 to specify $|\overline{\eta_s(d)B}|$ and $|\overline{\eta_s(d)C}|$. This can be done if we can first identify the corresponding location of $\eta_s(d)$ on $\overline{\mathbf{\Omega}}_s(d)$ as $d$ varies; the results are established in the next lemma.

**Lemma A.3:** Let $\eta_s(d)$ be defined in (2.27), and $\overline{\eta}_s(d)$ be the corresponding location on $\overline{\mathbf{\Omega}}_s(d)$. Also, let $\overline{U}_{l,s}(d)$ and $\overline{L}_{l,s}(d)$ be the points on $\overline{\mathbf{\Omega}}_s(d)$ that correspond to, respectively, $U_{l,s}(d)$ and $L_{l,s}(d)$ on $\mathbf{\Omega}_s(d)$. Denote by $\overline{E}(d)$ the point $(\sqrt{1-\delta} \times \frac{d_{\max}}{d}, \sqrt{1+\delta} \times \frac{d_{\max}}{d})$ on the coordinate system on Figure A.3. The following results hold:

$$\overline{\eta}_s(d) = \begin{cases} \overline{U}_{l,s}(d), & \text{if } (\tilde{d}_{\max}^2 + d^2)/2 < (1-\delta) + (1+\delta), \\ \overline{E}(d), & \text{if } (\tilde{d}_{\min}^2 + d^2)/2 \leq (1-\delta) + (1+\delta) \leq (\tilde{d}_{\max}^2 + d^2)/2, \\ \overline{L}_{l,s}(d), & \text{if } (\tilde{d}_{\min}^2 + d^2)/2 > (1-\delta) + (1+\delta). \end{cases} \quad (A.6)$$

*[Proof of Lemma A.3]:* We first note that

$$|\overline{BD}|\cos(\angle DBC) = |\overline{BD}| \times \frac{d_{\max}^2 + |\overline{BD}|^2 - |\overline{DC}|^2}{2|\overline{BD}| \times d_{\max}} = \frac{d_{\max}^2 + |\overline{BD}|^2 - |\overline{DC}|^2}{2 \times d_{\max}}, \quad (A.7)$$

which implies that $|\overline{BD}|\cos(\angle DBC)$ attains the minimum if $|\overline{BD}|^2 - |\overline{DC}|^2$ is minimized. Hence, $\eta_s(d)$ in (2.27) can be equivalently rewritten as $\eta_s(d) = \arg\min_{D\in\mathbf{\Omega}_s(d)}\{|\overline{BD}|^2 - |\overline{DC}|^2\}$, and thus

$$\overline{\eta}_s(d) = \arg\min_{D\in\overline{\mathbf{\Omega}}_s(d)}\{|\overline{BD}|^2 - |\overline{DC}|^2\} \quad (A.8)$$

To minimize $|\overline{BD}|^2 - |\overline{DC}|^2$, $|\overline{BD}|^2$ should be as small as possible, whereas $|\overline{DC}|^2$ should be instead as large as possible. It can be seen from Figure A.3 that the point $\overline{E}(d)$ is associated with the smallest $|\overline{BD}|^2$ and the largest $|\overline{DC}|^2$. Hence, among all points on $\overline{\mathbf{\Omega}}_s(d)$, the one located closest to $\overline{E}(d)$ will yield minimal $|\overline{BD}|^2 - |\overline{DC}|^2$, and therefore according to (A.8) should be identified as

$\overline{\eta}_s(d)$. Since $\overline{E}(d)$ is not always feasible (i.e., $\overline{E}(d)$ may lie outside $\overline{\Omega}_s(d)$), $\overline{\eta}_s(d)$ depends on the location of $\overline{E}(d)$ relative to $\overline{\Omega}_s(d)$. Under the assumption $(\tilde{d}_{\max}^2 + d^2)/2 < (1-\delta) + (1+\delta)$, $\overline{E}(d)$ locates above $\overline{\Omega}_s(d)$ (see Figure A.4-(a)), and the desired solution is thus $\overline{U}_{l,s}(d)$. In case that $(\tilde{d}_{\min}^2 + d^2)/2 \leq (1-\delta) + (1+\delta) \leq (\tilde{d}_{\max}^2 + d^2)/2$, $\overline{E}(d)$ is feasible (see Figure A.4-(b)) and thus $\overline{\eta}_s(d) = \overline{E}(d)$. If $(\tilde{d}_{\min}^2 + d^2)/2 > (1-\delta) + (1+\delta)$, $\overline{E}(d)$ locates on the left of $\overline{\Omega}_s(d)$ (see Figure A.4-(c)), and the desired solution is $\overline{\eta}_s(d) = \overline{L}_{l,s}(d)$.

*[Proof of Lemma 2.2]:* To prove (1), it suffices to show that, as $d$ increases from $d_{\min}$ to $d_{\max}$, the resultant $\cos\alpha$ is monotonically decreasing. Now, if $d$ is small such that $(\tilde{d}_{\max}^2 + d^2)/2 < (1-\delta) + (1+\delta)$, Lemma A.3 asserts that $\overline{\eta}_s(d) = \overline{U}_{l,s}(d)$ (cf. Figure A.4-(a)); since the magnitude triple $(|\overline{\eta_s(d)B}|, |\overline{\eta_s(d)C}|, |\overline{BC}|)$ associated with $\overline{\eta}_s(d) = \overline{U}_{l,s}(d)$ is given by $(\sqrt{1-\delta} \cdot (d_{\max}/d), \sqrt{(\tilde{d}_{\max}^2 + d^2)/2 - (1-\delta)} \cdot (d_{\max}/d), d \cdot (d_{\max}/d))$, the resultant $\alpha$ is easily determined by the law of cosines to be

$$\cos\alpha = \frac{\tilde{d}_{\max}^2 - d^2}{4\sqrt{(\tilde{d}_{\max}^2 + d^2)/2 - (1-\delta)}\sqrt{1-\delta}}. \tag{A.9}$$

As $d$ increases such that $(\tilde{d}_{\min}^2 + d^2)/2 \leq (1-\delta) + (1+\delta) \leq (\tilde{d}_{\max}^2 + d^2)/2$, we have from Lemma A.3 that $\overline{\eta}_s(d) = \overline{E}(d)$ (see Figure A.4-(b)); the associated magnitude triple is $(|\overline{\eta_s(d)B}|, |\overline{\eta_s(d)C}|, |\overline{BC}|)$ $= (\sqrt{1-\delta} \cdot (d_{\max}/d), \sqrt{1+\delta} \cdot (d_{\max}/d), d \cdot (d_{\max}/d))$, yielding

$$\cos\alpha = \frac{(1+\delta) + (1-\delta) - d^2}{2\sqrt{1+\delta}\sqrt{1-\delta}}. \tag{A.10}$$

As $d$ further increases such that $(\tilde{d}_{\min}^2 + d^2)/2 > (1-\delta) + (1+\delta)$, then $\overline{\eta}_s(d) = \overline{L}_{l,s}(d)$. Thus $(|\overline{\eta_s(d)B}|, |\overline{\eta_s(d)C}|, |\overline{BC}|) = (\sqrt{(\tilde{d}_{\min}^2 + d^2)/2 - (1+\delta)} \cdot (d_{\max}/d), \sqrt{1+\delta} \cdot (d_{\max}/d), d \cdot (d_{\max}/d))$, which results in

$$\cos\alpha = \frac{\tilde{d}_{\max}^2 - d^2}{4\sqrt{(\tilde{d}_{\min}^2 + d^2)/2 - (1+\delta)}\sqrt{1+\delta}}. \tag{A.11}$$

From (A.9)~(A.11), it can be seen that $\cos\alpha$ is monotonically decreasing with $d$. This thus proves the lemma. $\square$

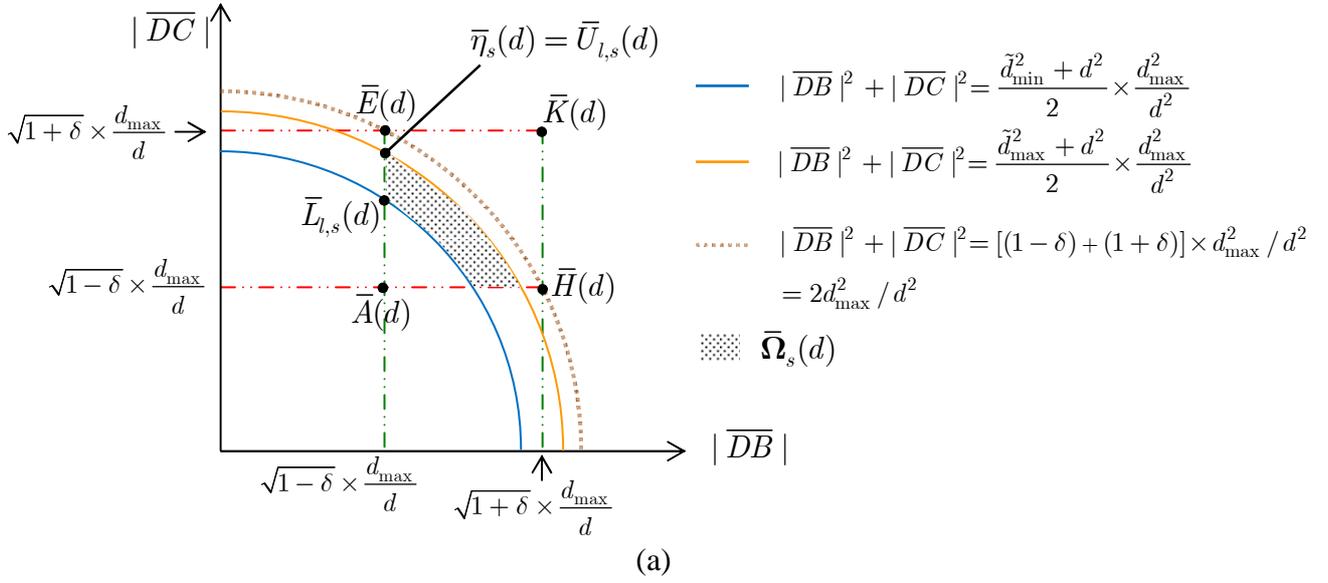

(a)

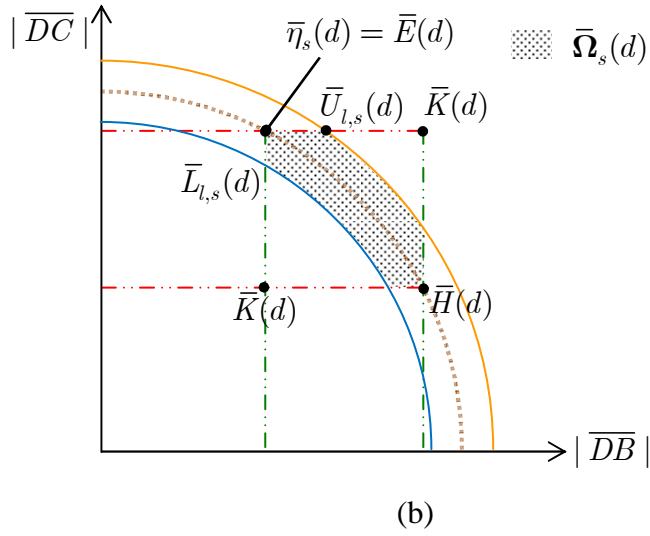

(b)

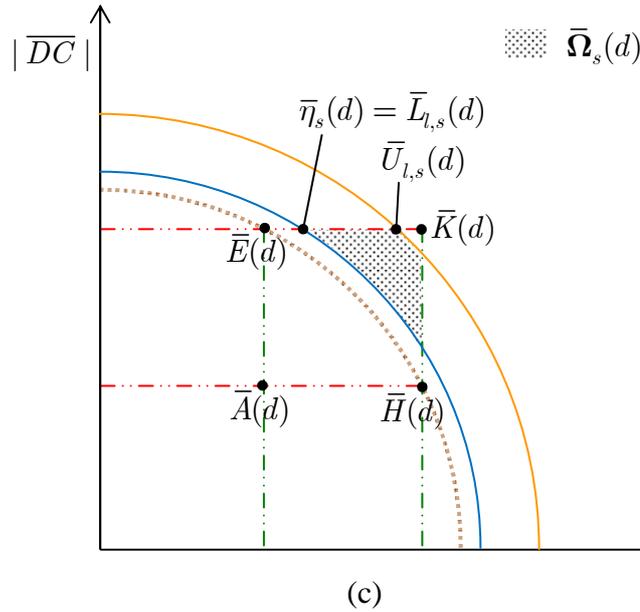

(c)

Figure A.4. Depiction of $\bar{\boldsymbol{\Omega}}_s(d)$ with respective to different $d$: (a) $(\tilde{d}_{\max}^2 + d^2)/2 < (1-\delta) + (1+\delta)$, (b) $(\tilde{d}_{\min}^2 + d^2)/2 \leq (1-\delta) + (1+\delta) \leq (\tilde{d}_{\max}^2 + d^2)/2$, (c) $(\tilde{d}_{\min}^2 + d^2)/2 > (1-\delta) + (1+\delta)$.

## B. Proof of Theorem 2.4

We shall first prove that $\alpha_{\max}$ is attained by some point on the bottom boundary $\partial\Omega_b$. For this let us construct a circle $\mathcal{C}_1$ (marked in purple) with $\overline{BC}$ as a chord and with $F$ and $G$ on arc $BC$ as depicted in Figure A.5-(a), shown on the next page; also, let

$$\Gamma_1 \triangleq \{D \in \Omega \mid D \text{ lies on or inside } \mathcal{C}_1\}, \tag{A.12}$$

and denote by $\Gamma_1^c$ the complement of $\Gamma_1$ in $\Omega$. By Lemma A.2, it follows immediately that $\angle BDC < \angle BGC$ for all $D \in \Gamma_1^c$. Notably, all the points on the left and right boundary, i.e. $\partial\Omega_l$ and $\partial\Omega_r$, expect $F$ and $G$, must belong to $\Gamma_1^c$; this is because if there exists some $D' \in \partial\Omega_r \setminus \{G\}$ in $\Gamma_1$, then again by Lemma A.2 we must have $\angle BD'C \geq \angle BGC$, which contradicts with part (2) of Lemma 2.2. As a result, we can rule out those $D \in \Gamma_1^c$ regarding the search of $\alpha_{\max}$. It then remains to consider $\Gamma_1$. Observe that, for any $D \in \Gamma_1$, both the two sides $\overline{DB}$ and $\overline{DC}$ of $\triangle DBC$ intersect with $\partial\Omega_b$, meaning that there exists an $A \in \partial\Omega_b$ such that $\triangle ABC$ lies entirely inside $\triangle DBC$ (see also Figure 5-(a)). By Lemma A.1, we have $\angle BAC > \angle BDC$. This shows that $\alpha_{\max}$ must be attained by some point on $\partial\Omega_b$.

Now let's go on to show that $\alpha_{\min}$ is achieved by some point $D \in \partial\Omega_t$. The idea is quite similar to that in the previous proof. Firstly, we construct circles $\mathcal{C}_2$ with $\overline{BC}$ as a chord and with $P$ and $Q$ on arc $BC$ (see Figure A.5-(b)). Similar to (A.12) we define

$$\Gamma_2 \triangleq \{D \in \Omega \mid D \text{ lies on or inside } \mathcal{C}_2\}; \tag{A.13}$$

also, we denote by $\Gamma_2^c$ the complement of $\Gamma_2$ in $\Omega$. From Lemma A.2, we immediately have $\angle BD'C > \angle BPC$ for all $D' \in \Gamma_2$. In addition, from Lemma A.2 and Lemma 2.2, $\partial\Omega_l \setminus \{P\}$ and $\partial\Omega_r \setminus \{Q\}$ both must belong to $\Gamma_2$. Hence, it only remains to consider $\Gamma_2^c$ regarding the identification of $\alpha_{\min}$. Observe that, for any $D' \in \Gamma_2^c$, there exists at least one element $A'$ on $\partial\Omega_t$ such that $\triangle D'BC$ is completely inside $\triangle A'BC$ (see also Figure 5-(b)). From Lemma A.1, we must have $\angle BA'C < \angle BD'C$. This established that $\alpha_{\min} = \angle BA'C$ for some point $A' \in \partial\Omega_t$. □

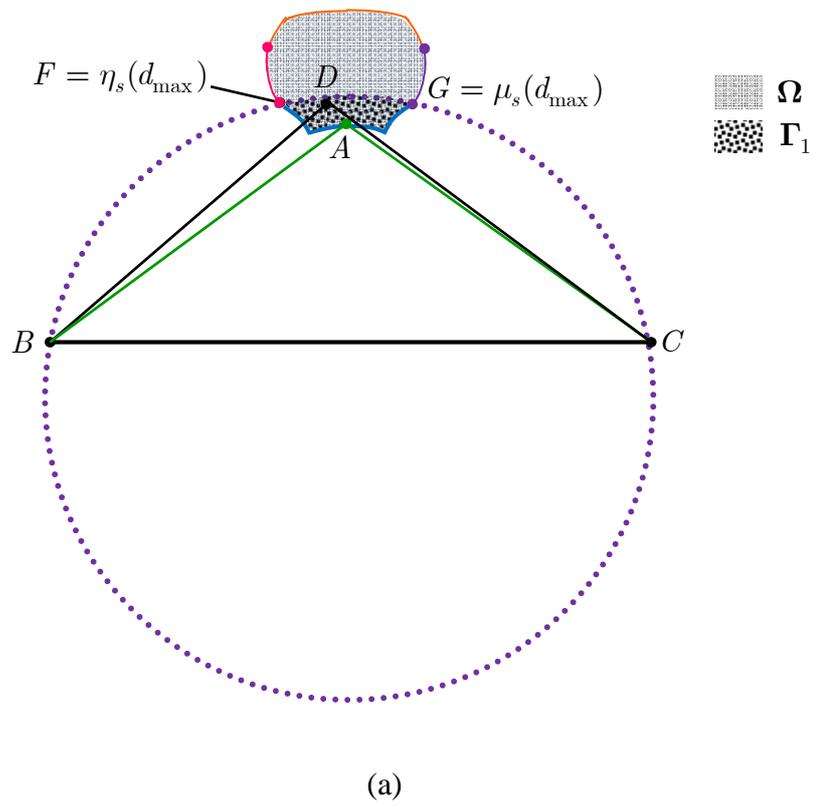

(a)

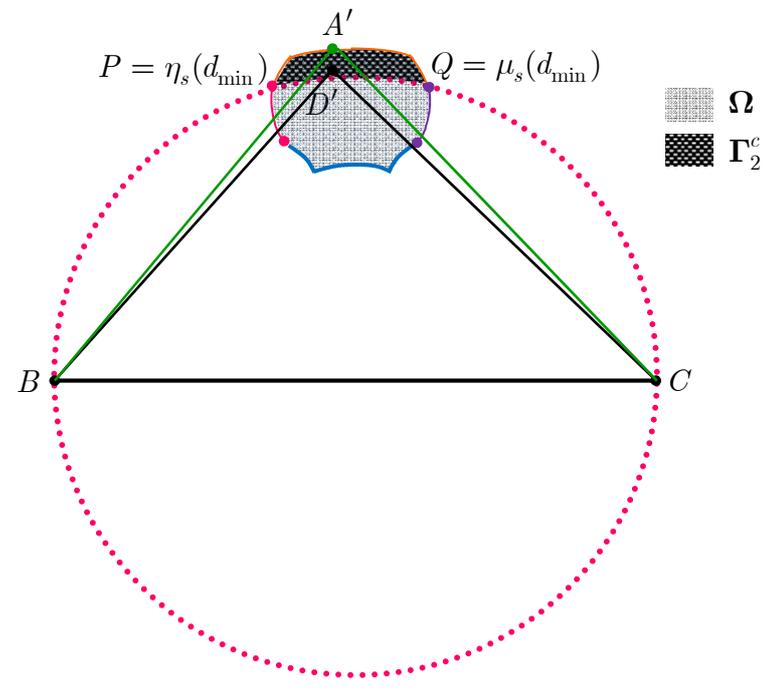

(b)

Figure A.5. Depiction of all geometric objects for plane geometric analyses conducted in the proof Theorem 2.4.

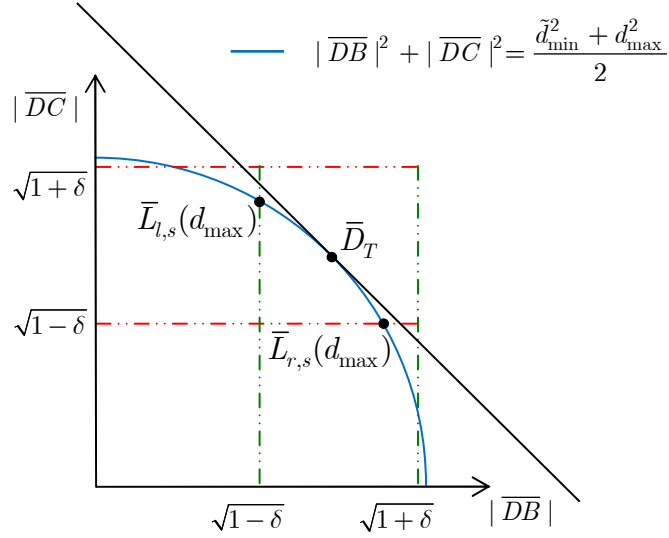

Figure A.6. Location of the tangency point $\bar{D}_T$ with respect to the curve $\mathcal{CV}(\bar{L}_l(d_{\max}), \bar{L}_r(d_{\max}))$.

## C. Proof of Lemma 3.1

Since the feasible set for Problem (P1) consists of merely these $D \in \mathcal{CV}(L_{l,s}(d_{\max}), L_{r,s}(d_{\max}))$, let us focus on the corresponding curve $\mathcal{CV}(\bar{L}_{l,s}(d_{\max}), \bar{L}_{r,s}(d_{\max})) \subset \bar{\mathbf{\Omega}}_s(d_{\max})$, which lies on the quarter-circle with radius $\sqrt{(\tilde{d}_{\min}^2 + d_{\max}^2)/2}$ as depicted in Figure A.6. It is clear that the maximal $|\overline{DB}| + |\overline{DC}|$ on the quarter-circle is attained by the unique tangency point, say, $\bar{D}_T$, of the tangent line $|\overline{DB}| + |\overline{DC}| = \sqrt{\tilde{d}_{\min}^2 + d_{\max}^2}$. Note that the coordinate of $\bar{D}_T$ is $(|\overline{DB}|, |\overline{DC}|) = (\sqrt{(\tilde{d}_{\min}^2 + d_{\max}^2)/4}, \sqrt{(\tilde{d}_{\min}^2 + d_{\max}^2)/4})$, and thus $\bar{D}_T$ exactly corresponds to $D_L$ defined in (3.4). As a result, the point $D_L$ will yield the maximal $\angle BDC$. The proof is thus completed. □

## D. Proof of Theorem 3.2

First of all, we note from Lemma 3.1 that

$$\angle BD_L C > \angle BDC \quad \text{for all} \quad D \in \mathcal{CV}(L_{l,s}(d_{\max}), L_{r,s}(d_{\max})) \setminus \{D_L\}. \tag{A.14}$$

Let us then construct a circle $\mathcal{C}_{D_L}$, marked in dashed blue in Figure A.7, with $\overline{BC}$ as a chord and $D_L$ on arc $BC$. From (A.14) and Lemma A.2, we must have $\mathcal{C}_{D_L} \cap \mathcal{CV}(L_{l,s}(d_{\max}), L_{r,s}(d_{\max})) = \{D_L\}$, and, in particular, $L_{l,s}(d_{\max})$ and $L_{r,s}(d_{\max})$ must locate outside $\mathcal{C}_{D_L}$. We claim that both $\mathcal{CV}(F, L_{l,s}(d_{\max}))$ and $\mathcal{CV}(L_{r,s}(d_{\max}), G)$ are also outside $\mathcal{C}_{D_L}$. As a result, it follows again from Lemma A.2 that

$$\angle BD_L C > \angle BDC \quad \text{for all} \quad D \in \mathcal{CV}(F, L_{l,s}(d_{\max})) \cup \mathcal{CV}(L_{r,s}(d_{\max}), G). \tag{A.15}$$

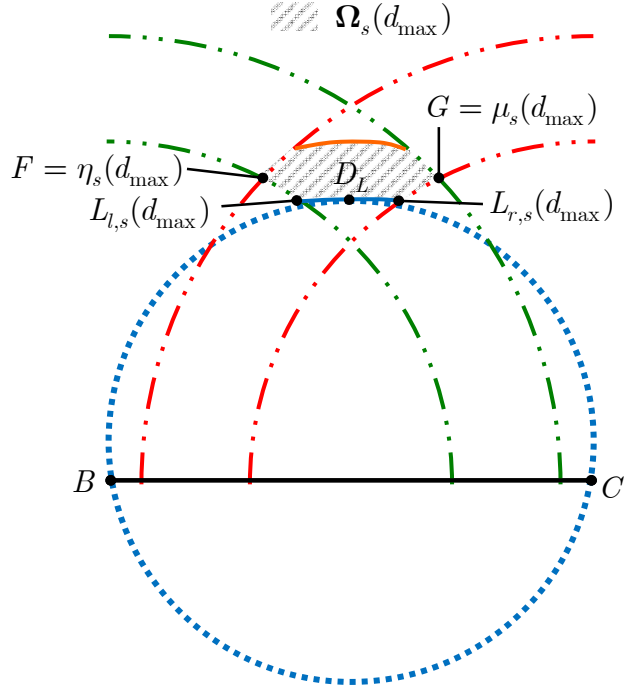

Figure A.7. Depiction of all geometric objects for plane geometric analyses conducted in the proof Theorem 3.2.

Based on (A.14) and (A.15) it can be deduced that $\alpha_{\max} = \angle BD_L C$. Since $(|\overline{BD_L}|, |\overline{D_L C}|, |\overline{BC}|) = (\sqrt{(\tilde{d}_{\min}^2 + d_{\max}^2)/4}, \sqrt{(\tilde{d}_{\min}^2 + d_{\max}^2)/4}, d_{\max})$, equation (3.5) follows. To prove the claim, due to the symmetric nature of the figure it suffices to show $\mathcal{CV}(F, L_{l,s}(d_{\max}))$ is outside $\mathcal{C}_{D_L}$. For this we need the next lemma.

**Lemma A.4:** Assume that $\mathcal{CV}(F, L_{l,s}(d_{\max}))$ contains more than one element. Then for all $D \in \mathcal{CV}(F, L_{l,s}(d_{\max}))$, we have $|\overline{DB}| = \sqrt{1-\delta}$.

*[Proof]:* The assertion of the lemma can be obtained by resorting to $\bar{\Omega}_s(d_{\max})$ shown in Figure A.4. To prove the lemma, let us first identify the region in $\bar{\Omega}_s(d_{\max})$ that corresponds to $\mathcal{CV}(F, L_{l,s}(d_{\max}))$. Denote by $\bar{F} \in \bar{\Omega}_s(d_{\max})$ the point corresponding to $F \in \Omega_s(d_{\max})$. By definition, $F = \eta_s(d_{\max})$ and, thus, $\bar{F} = \bar{\eta}_s(d_{\max})$. According to Lemma A.3, we then have

$$\bar{F} = \bar{\eta}_s(d_{\max}) = \begin{cases} \bar{U}_{l,s}(d_{\max}), & \text{if } (\tilde{d}_{\max}^2 + d_{\max}^2)/2 < (1-\delta) + (1+\delta), \\ \bar{E}(d_{\max}), & \text{if } (\tilde{d}_{\min}^2 + d_{\max}^2)/2 \leq (1-\delta) + (1+\delta) \leq (\tilde{d}_{\max}^2 + d_{\max}^2)/2, \\ \bar{L}_{l,s}(d_{\max}), & \text{if } (\tilde{d}_{\min}^2 + d_{\max}^2)/2 > (1-\delta) + (1+\delta). \end{cases} \quad (A.16)$$

which implies that

$$\mathcal{CV}(\bar{F}, \bar{L}_{l,s}(d_{\max})) = \begin{cases} \mathcal{CV}(\bar{U}_{l,s}(d_{\max}), \bar{L}_{l,s}(d_{\max})), & \text{if } (\tilde{d}_{\max}^2 + d_{\max}^2)/2 < (1-\delta) + (1+\delta), \\ \mathcal{CV}(\bar{E}(d_{\max}), \bar{L}_{l,s}(d_{\max})), & \text{if } (\tilde{d}_{\min}^2 + d_{\max}^2)/2 \leq (1-\delta) + (1+\delta) \leq (\tilde{d}_{\max}^2 + d_{\max}^2)/2, \\ \bar{L}_{l,s}(d_{\max}), & \text{if } (\tilde{d}_{\min}^2 + d_{\max}^2)/2 > (1-\delta) + (1+\delta). \end{cases}$$

(A.17)

Under the assumption that $\mathcal{CV}(F, L_{l,s}(d_{\max}))$ contains more than one element and from (A.17), the curve on $\overline{\Omega}_s(d_{\max})$ corresponding to $\mathcal{CV}(F, L_{l,s}(d_{\max}))$ is either $\mathcal{CV}(\overline{U}_{l,s}(d_{\max}), \overline{L}_{l,s}(d_{\max}))$ or $\mathcal{CV}(\overline{E}(d_{\max}), \overline{L}_{l,s}(d_{\max}))$ (this is depicted, respectively, in Figures A.4-(a) and -(b)). We can observe from the figures that $\mathcal{CV}(\overline{U}_{l,s}(d_{\max}), \overline{L}_{l,s}(d_{\max}))$ and $\mathcal{CV}(\overline{E}(d_{\max}), \overline{L}_{l,s}(d_{\max}))$ both locate on the line segment $\overline{A(d_{\max})\overline{E}(d_{\max})}$. The assertion follows since all points on $\overline{A(d_{\max})\overline{E}(d_{\max})}$ are associated with $|\overline{DB}| = \sqrt{1-\delta}$. □

*[Proof of Claim]:* If $\mathcal{CV}(F, L_{l,s}(d_{\max}))$ degenerates into one point $\{L_{l,s}(d_{\max})\}$, we are done since $L_{l,s}(d_{\max})$ is outside $\mathcal{C}_{D_L}$. For the general case, it readily follows from Lemma A.4 that $\mathcal{CV}(F, L_{l,s}(d_{\max}))$ lies on the arc of $\mathcal{C}(B, \sqrt{1-\delta})$. Since $L_{l,s}(d_{\max})$ is outside $\mathcal{C}_{D_L}$ and, moreover, $F = \eta_s(d_{\max})$ locates to the left of $L_{l,s}(d_{\max})$ (see Figure A.7), $\mathcal{CV}(F, L_{l,s}(d_{\max})) \subset \mathcal{C}(B, \sqrt{1-\delta})$ must lie entirely outside $\mathcal{C}_{D_L}$. □

## E. Proof of Lemma 3.3

Now the problem is to minimize $|\overline{DB}| + |\overline{DC}|$ under the same constraints as in problem (P1). The solution can be found by again resorting to Figure A.6 used in the proof of Lemma 3.1. It is easy to see from the figure that, to minimize $|\overline{DB}| + |\overline{DC}|$ among all $D \in \mathcal{CV}(\overline{L}_{l,s}(d_{\max}), \overline{L}_{r,s}(d_{\max}))$, we shall choose the one that is located as far away from the tangency point $\overline{D}_T$ as possible. The solution is either one of the two end points of $\mathcal{CV}(\overline{L}_{l,s}(d_{\max}), \overline{L}_{r,s}(d_{\max}))$, namely, $\overline{L}_{l,s}(d_{\max})$ or $\overline{L}_{r,s}(d_{\max})$. The corresponding point on $\Omega_s(d_{\max})$ is $D = L_{l,s}(d_{\max})$ or $D = L_{r,s}(d_{\max})$. The assertion thus follows. □

## F. Proof of Theorem 3.4

Now if we construct a circle $C_3$ with $\overline{BC}$ as a chord and with $L_{l,s}(d_{\max})$ and $L_{r,s}(d_{\max})$ on arc $BC$ as in Figure A.8 shown on the next page, by following similar procedures as in the proof of Theorem 3.2 it can be verified that $\mathcal{CV}(L_{l,s}(d_{\max}), L_{r,s}(d_{\max})) \setminus \{L_{l,s}(d_{\max}), L_{r,s}(d_{\max})\}$, $\mathcal{CV}(F, L_{l,s}(d_{\max}))$, and $\mathcal{CV}(L_{r,s}(d_{\max}), G)$ all lie outside this circle. That is, the whole region $\Omega_s(d_{\max})$, except $L_{l,s}(d_{\max})$ and $L_{r,s}(d_{\max})$, falls outside circle $C_3$. Again from Lemma A.2, we thus have

$$\alpha_{\max} = \angle BL_{r,s}(d_{\max})C = \angle BL_{l,s}(d_{\max})C = \cos^{-1}\left(\frac{|\overline{L_{l,s}(d_{\max})B}|^2 + |\overline{L_{l,s}(d_{\max})C}|^2 - d_{\max}^2}{2|\overline{L_{l,s}(d_{\max})B}| \times |\overline{L_{l,s}(d_{\max})C}|}\right). \quad (A.18)$$

To specify $|\overline{L_{l,s}(d_{\max})B}|$ and $|\overline{L_{l,s}(d_{\max})C}|$, we can first refer to $\overline{\Omega}(d_{\max})$ in Figure A.4 to determine the location of $\overline{L}_{l,s}(d_{\max})$ which corresponds to $L_{l,s}(d_{\max})$; this in turn allows us to determine $|\overline{L_{l,s}(d_{\max})B}|$ and $|\overline{L_{l,s}(d_{\max})C}|$. Under the assumption $(\tilde{d}_{\min}^2 + d_{\max}^2)/2 \leq (1-\delta) + (1+\delta)$, i.e., $(\tilde{d}_{\min}^2 + d_{\max}^2) \leq 4$, we have $|\overline{L_{l,s}(d_{\max})B}| = \sqrt{1-\delta}$ (see Figures A.4-(a) and -(b)), and by computation $|\overline{L_{l,s}(d_{\max})C}| = \sqrt{(\tilde{d}_{\min}^2 + d_{\max}^2)/2 - (1-\delta)}$ (since $|\overline{L_{l,s}(d_{\max})B}|^2 + |\overline{L_{l,s}(d_{\max})C}|^2 = (\tilde{d}_{\min}^2 + d_{\max}^2)/2$), which together with (A.18) leads to (3.6). In case that $(\tilde{d}_{\min}^2 + d_{\max}^2)/2 > (1-\delta) + (1+\delta)$, i.e., $(\tilde{d}_{\min}^2 + d_{\max}^2) > 4$, we have $|\overline{L_{l,s}(d_{\max})C}| = \sqrt{1+\delta}$, and $|\overline{L_{l,s}(d_{\max})B}| = \sqrt{(\tilde{d}_{\min}^2 + d_{\max}^2)/2 - (1+\delta)}$ (since $|\overline{L_{l,s}(d_{\max})B}|^2 + |\overline{L_{l,s}(d_{\max})C}|^2 = (\tilde{d}_{\min}^2 + d_{\max}^2)/2$), which combined with (A.18) yields (3.7). We note that, since $\tilde{d}_{\min}^2 - d_{\max}^2 < 0$ and from (3.2), we must have $\cos(\angle BL_{l,s}(d_{\max})C) < 0$ in the considered scenario. In case that

$$\frac{|\overline{L_{l,s}(d_{\max})B}|^2 + |\overline{L_{l,s}(d_{\max})C}|^2 - d_{\max}^2}{2|\overline{L_{l,s}(d_{\max})B}| \times |\overline{L_{l,s}(d_{\max})C}|} < -1, \tag{A.19}$$

we must have $\cos\alpha_{\max} = -1$, and hence $\alpha_{\max} = \pi$. $\square$

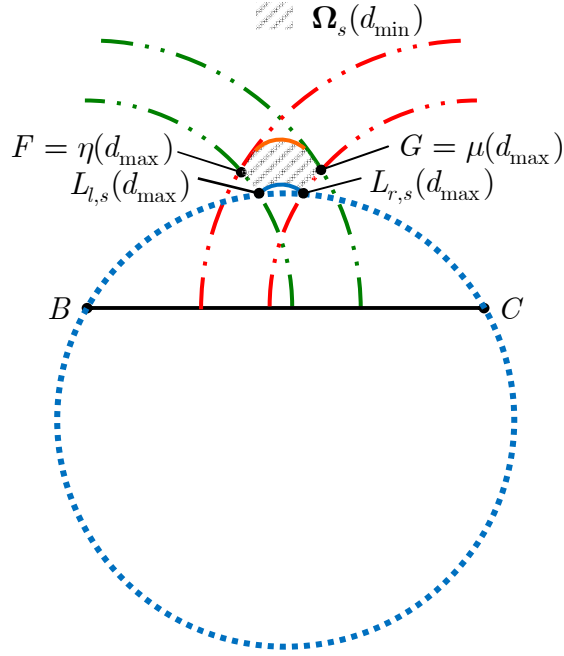

Figure A.8 Depiction of all geometric objects for plane geometric analyses conducted in the proof Theorem 3.4.

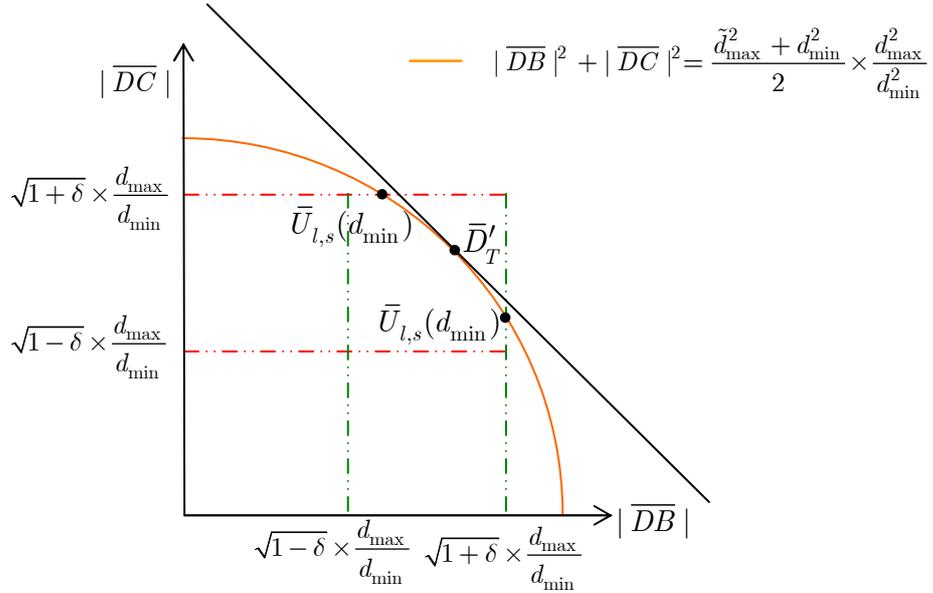

Figure A.9. Location of the tangency point $\bar{D}'_T$ with respect to the curve $\mathcal{CV}(\bar{U}_{l,s}(d_{\max})\bar{U}_{r,s}(d_{\max}))$.

### G. Proof of Lemma 4.1

The idea behind this proof is essentially the same as that in the proof of Lemma 3.1. Since the feasible set associated with (P3) consists of all $D \in \mathcal{CV}(U_{l,s}(d_{\min}), U_{r,s}(d_{\min}))$, let us focus on the corresponding curve $\mathcal{CV}(\bar{U}_{l,s}(d_{\min}), \bar{U}_{r,s}(d_{\min})) \subset \bar{\mathbf{\Omega}}_s(d_{\min})$, which lies on the quarter-circle with radius $\sqrt{(\tilde{d}_{\max}^2 + d_{\min}^2)/2} \cdot \xi$. Denote by $\bar{D}'_T$ the unique tangency point of the tangent line $|\overline{DB}| + |\overline{DC}| = \sqrt{\tilde{d}_{\max}^2 + d_{\min}^2} \cdot \xi$. Note that the coordinate of $\bar{D}'_T$ is $(|\overline{DB}|, |\overline{DC}|) = (\sqrt{(\tilde{d}_{\max}^2 + d_{\min}^2)/4} \cdot \xi, \sqrt{(\tilde{d}_{\max}^2 + d_{\min}^2)/4} \cdot \xi)$, and thus $\bar{D}'_T$ exactly corresponds to $D_U$ defined in (4.4). As a result, the point $D_U$ will yield the minimal $\angle BDC$. The proof is thus completed. $\square$

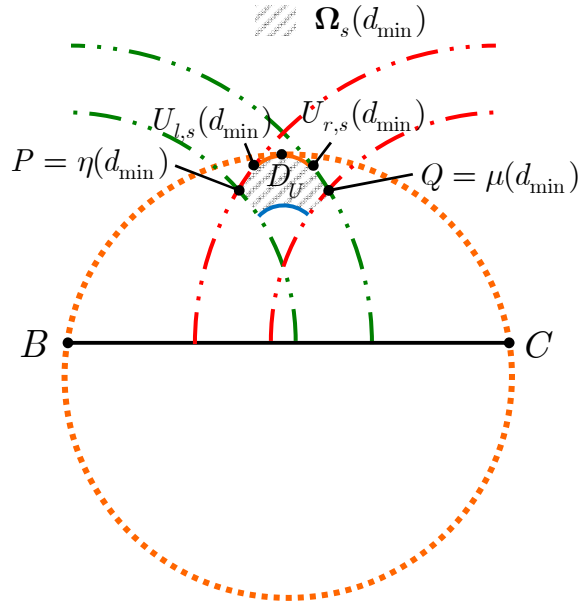

Figure A.10. Depiction of all geometric objects for plane geometric analyses conducted in the proof Theorem 4.2.

### H. Proof of Theorem 4.2

The idea behind the proof is quite similar to that in the proof of Theorem 3.2. From Lemma 4.1 it follows immediately

$$\angle BD_U C < \angle BDC \text{ for all } D \in \mathcal{CV}(U_{l,s}(d_{\min}), U_{r,s}(d_{\min})) \setminus \{D_U\}. \tag{A.20}$$

Construct a circle $\mathcal{C}_{D_U}$ with $\overline{BC}$ as a chord and with $D_U$ on arc $BC$ (see Figure A.10). Then $\mathcal{CV}(U_{l,s}(d_{\min}), U_{r,s}(d_{\min})) \setminus \{D_U\}$ must lie entirely within $\mathcal{C}_{D_U}$; if there exists some $D' \in \mathcal{CV}(U_{l,s}(d_{\min}), U_{r,s}(d_{\min})) \setminus \{D_U\}$ on or outside $\mathcal{C}_{D_U}$, then Lemma A.2 implies $\angle BD_U C \geq \angle BD'C$, which contradicts with (A.20). We further claim that both $\mathcal{CV}(P, U_{l,s}(d_{\min}))$ and $\mathcal{CV}(U_{r,s}(d_{\min}), Q)$ are also inside $\mathcal{C}_{D_U}$. Hence, again by Lemma A.2 we have

$$\angle BD_U C < \angle BDC \quad \text{for all} \quad D \in \mathcal{CV}(P, U_{l,s}(d_{\min})) \cup \mathcal{CV}(U_{r,s}(d_{\min}), Q). \tag{A.21}$$

Based on (A.20) and (A.21), it can be deduced that $\alpha_{\min} = \angle BD_U C$. To prove the claim, due to the symmetric nature of the figure it suffices to show $\mathcal{CV}(P, U_{l,s}(d_{\min}))$ is inside $\mathcal{C}_{D_U}$. This can be done with the aid of the next lemma.

***Lemma A.5:*** Assume that $\mathcal{CV}(P, U_{l,s}(d_{\min}))$ contains more than one element. For all $D \in \mathcal{CV}(P, U_{l,s}(d_{\min}))$, we have $|\overline{DC}| = \sqrt{1+\delta} \cdot \xi$.

*[Proof]:* The assertion of the lemma can be obtained by resorting to $\overline{\boldsymbol{\Omega}}_s(d_{\min})$ shown in Figure A.3. Let us first identify the curve in $\overline{\boldsymbol{\Omega}}_s(d_{\min})$ that corresponds to $\mathcal{CV}(P, U_{l,s}(d_{\min})) \subset \boldsymbol{\Omega}_s(d_{\min})$. By definition, $P = \eta_s(d_{\min})$, the left end point of $\partial \boldsymbol{\Omega}_s^+(d_{\min})$, and thus the corresponding point on $\overline{\boldsymbol{\Omega}}_s(d_{\min})$ is $\overline{P} = \overline{\eta}_s(d_{\min})$. Note that the exact location of $\overline{P} = \overline{\eta}_s(d_{\min})$ as a function of $d$ can be determined by (A.6) with $d = d_{\min}$. This in turn implies

$$\mathcal{CV}(\overline{P}, \overline{U}_{l,s}(d_{\min})) = \begin{cases} \overline{U}_{l,s}(d_{\min}), & (\tilde{d}_{\max}^2 + d_{\min}^2)/2 < (1-\delta) + (1+\delta), \\ \mathcal{CV}(\overline{E}(d_{\min}), \overline{U}_{l,s}(d_{\min})), & \text{if } (\tilde{d}_{\min}^2 + d_{\min}^2)/2 \leq (1-\delta) + (1+\delta) \leq (\tilde{d}_{\max}^2 + d_{\min}^2)/2, \\ \mathcal{CV}(\overline{L}_{l,s}(d_{\min}), \overline{U}_{l,s}(d_{\min})), & \text{if } (\tilde{d}_{\min}^2 + d_{\min}^2)/2 > (1-\delta) + (1+\delta). \end{cases} \tag{A.22}$$

Under the assumption that $\mathcal{CV}(\overline{P}, \overline{U}_{l,s}(d_{\min}))$ contains more than one element and from (A.22), the exact location of $\mathcal{CV}(\overline{P}, \overline{U}_{l,s}(d_{\min}))$ on $\overline{\boldsymbol{\Omega}}(d_{\min})$ is either $\mathcal{CV}(\overline{E}(d_{\min}), \overline{U}_{l,s}(d_{\min}))$ or $\mathcal{CV}(\overline{L}_{l,s}(d_{\min}), \overline{U}_{l,s}(d_{\min}))$ (this is depicted, respectively, in Figures A.4-(b) and -(c)). It can be easily seen from the figures that, for either case, $\mathcal{CV}(\overline{P}, \overline{U}_{l,s}(d_{\min}))$ is located on the line segment $\overline{\overline{E}(d_{\min})\overline{K}(d_{\min})}$. The assertion follows since all points on $\overline{\overline{E}(d_{\min})\overline{K}(d_{\min})}$ are associated with $|\overline{DC}| = \sqrt{1+\delta} \cdot \xi$. $\square$

*[Proof of Claim]:* If $\mathcal{CV}(P, U_{l,s}(d_{\min}))$ degenerates into one point $\{U_l\}$, we are done since $U_{l,s}(d_{\min})$ is inside $\mathcal{C}_{D_U}$. For the general case, it readily follows from Lemma A.5 that $\mathcal{CV}(P, U_{l,s}(d_{\min}))$ lies on the arc of $\mathcal{C}(C, \sqrt{1+\delta} \cdot \xi)$. Since $U_{l,s}(d_{\min})$ is inside $\mathcal{C}_{D_U}$, and, moreover, $P = \eta_s(d_{\min})$ locates to the left of $U_{l,s}(d_{\min})$ (see Figure A.11), $\mathcal{CV}(P, U_{l,s}(d_{\min})) \subset \mathcal{C}(C, \sqrt{1+\delta} \cdot \xi)$ must lie entirely inside $\mathcal{C}_{D_U}$.

$\square$

*I. Proof of Lemma 4.3*

The proof can be done based on Lemma 4.1, and by following the arguments as in the proof of Lemma 3.3. $\square$

*J. Proof of Theorem 4.4*

From Lemma 4.3, we have

$$\angle BDC > \angle BU_{l,s}(d_{\min})C = \angle BU_{r,s}(d_{\min})C \text{ for all } D \in \mathcal{CV}(U_{l,s}(d_{\min}), U_{r,s}(d_{\min})) \setminus \{U_{l,s}(d_{\min}), U_{r,s}(d_{\min})\} \tag{A.23}$$

It then suffices to consider $\mathcal{CV}(P, U_{l,s}(d_{\min}))$ for identifying $\alpha_{\min}$ since the region $\Omega$ is symmetric. If $\mathcal{CV}(P, U_{l,s}(d_{\min}))$ degenerates into one single element $\{U_{l,s}(d_{\min})\}$ (by Lemma A.5, this case occurs when $(\tilde{d}_{\max}^2 + d_{\min}^2)/2 < (1+\delta) + (1-\delta)$, i.e., $(\tilde{d}_{\max}^2 + d_{\min}^2) < 4$), then $\angle BU_{l,s}(d_{\min})C$ is the desired solution; the magnitude triple associated with $U_{l,s}(d_{\min})$ is $(|\overline{BU_{l,s}(d_{\min})}|, |\overline{CU_{l,s}(d_{\min})}|, |\overline{BC}|) = (\sqrt{1-\delta}\cdot\xi, \sqrt{(\tilde{d}_{\max}^2 + d_{\min}^2)/2 - (1-\delta)}\cdot\xi, d_{\min}\cdot\xi)$, resulting in

$$\cos\alpha_{\min} = \frac{(\tilde{d}_{\max}^2 - d_{\min}^2)}{4\sqrt{1-\delta}\sqrt{(\tilde{d}_{\max}^2 + d_{\min}^2)/2 - (1-\delta)}}, \tag{A.24}$$

which yields (4.6). Let's turn to consider the general case when $\mathcal{CV}(P, U_{l,s}(d_{\min}))$ contains more than one element (the condition under which this occurs is $(\tilde{d}_{\max}^2 + d_{\min}^2)/2 \geq (1+\delta) + (1-\delta)$, i.e., $(\tilde{d}_{\max}^2 + d_{\min}^2) \geq 4$, as can be seen (A.22)). From Lemma A.5, it follows that $|\overline{DC}| = \sqrt{1+\delta}\cdot\xi$ for all $D \in \mathcal{CV}(P, U_{l,s}(d_{\min}))$. Since $|\overline{BC}| = d_{\max} = d_{\min}\cdot\xi$, the cosine of $\angle BDC$ for $D \in \mathcal{CV}(P, U_{l,s}(d_{\min}))$ reads

$$\begin{aligned}\cos(\angle BDC) &= \frac{|\overline{DB}|^2 + |\overline{DC}|^2 - |\overline{BC}|^2}{2|\overline{DB}|\times|\overline{DC}|} \\ &= \frac{|\overline{DB}|^2 + (1+\delta)\cdot\xi^2 - d_{\min}^2\cdot\xi^2}{|\overline{DB}|} \times \gamma_2, \text{ where } \gamma_2 \triangleq \frac{1}{2\sqrt{1+\delta}\xi} > 0.\end{aligned} \tag{A.25}$$

Equation (A.25) shows that, for $D \in \mathcal{CV}(P, U_{l,s}(d_{\min}))$, the resultant $\cos(\angle BDC)$ is completely determined by $|\overline{DB}|$. Hence, to find the minimal $\angle BDC$, it is equivalent to find the optimal $|\overline{DB}|$ which maximizes $\cos(\angle BDC)$. By treating $|\overline{DB}|$ as a dummy variable, let us first compute the first- and second-order derivatives of $\cos(\angle BDC)$ with respect to $|\overline{DB}|$ as

$$(\cos(\angle BDC))' = \left(1 - \frac{(1+\delta)\cdot\xi^2 - d_{\min}^2\cdot\xi^2}{|\overline{DB}|^2}\right) \times \gamma_2, \tag{A.26}$$

and

$$(\cos(\angle BDC))'' = 2((1+\delta) - d_{\min}^2) \times \xi^2 \times \gamma_2 / |\overline{DB}|^3. \tag{A.27}$$

Let us first consider the case $(1+\delta) - d_{\min}^2 \geq 0$, which together with (A.27) asserts that $\cos(\angle BDC)$ is a convex function of $|\overline{DB}|$. The convexity of $\cos(\angle BDC)$ implies that the maximal $\cos(\angle BDC)$ (or, minimal achievable $\angle BDC$) is attained by either the minimal $|\overline{DB}|$ or the maximal $|\overline{DB}|$ among all $D \in \mathcal{CV}(P, U_{l,s}(d_{\min}))$. Let us refer to Figures A.4-(b) and A.4-(c) with $d = d_{\min}$, and focus on the corresponding curve $\mathcal{CV}(\overline{P},\overline{U}_{l,s}(d_{\min}))$. It is easy to see that the minimal and maximal $|\overline{DB}|$ are attained by, respectively, $\overline{P}$ and $\overline{U}_{l,s}(d_{\min})$. This thus implies that the minimal and maximal $|\overline{DB}|$ among all $D \in \mathcal{CV}(P, U_{l,s}(d_{\min}))$ are attained by either one of the two end points, namely $D = P$ and $D = U_{l,s}(d_{\min})$; as a result, we have

$$\cos \alpha_{\min} = \max\{\cos \angle BPC, \cos \angle BU_{l,s}(d_{\min})C\}. \tag{A.28}$$

To compute $\alpha_{\min}$ based on (A.28), we shall first find explicit expressions for $\cos \angle BPC$ and $\cos \angle BU_{l,s}(d_{\min})C$. Since $(\tilde{d}_{\max}^2 + d_{\min}^2)/2 \geq (1+\delta) + (1-\delta)$, we can refer to Figure A.4-(b) and -(c) to determine the magnitude triple associated with $U_{l,s}(d_{\min})$ as $(|\overline{U_{l,s}(d_{\min})B}|, |\overline{U_{l,s}(d_{\min})C}|, |\overline{BC}|)$
$= (\sqrt{(\tilde{d}_{\max}^2 + d_{\min}^2)/2 - (1+\delta)} \cdot \xi, \sqrt{1+\delta} \cdot \xi, d_{\min} \cdot \xi)$; by the law of cosines it follows

$$\cos(\angle BU_{l,s}(d_{\min})C) = \frac{(\tilde{d}_{\max}^2 - d_{\min}^2)}{4\sqrt{1+\delta}\sqrt{(\tilde{d}_{\max}^2 + d_{\min}^2)/2 - (1+\delta)}}. \tag{A.29}$$

To find $\angle BPC$, we first note that the location corresponding to $P$ on the diagram $\overline{\boldsymbol{\Omega}}_s(d_{\min})$ is $\overline{\eta}(d_{\min})$. From Lemma A.3, the exact location of $\overline{\eta}(d_{\min})$ depends on the value of $d_{\min}$. If $(\tilde{d}_{\min}^2 + d_{\min}^2)/2 > (1+\delta) + (1-\delta)$, then $\overline{\eta}(d_{\min}) = \overline{L}_{l,s}(d_{\min})$, and thus $P = L_{l,s}(d_{\min})$. The associated magnitude triple is $(|\overline{PB}|, |\overline{PC}|, |\overline{BC}|) = (\sqrt{(\tilde{d}_{\min}^2 + d_{\min}^2)/2 - (1+\delta)} \cdot \xi, \sqrt{1+\delta} \cdot \xi, d_{\min} \cdot \xi)$, which implies

$$\cos(\angle BPC) = \cos(\angle BL_{l,s}(d_{\min})C) = \frac{(\tilde{d}_{\min}^2 - d_{\min}^2)}{4\sqrt{1+\delta}\sqrt{(\tilde{d}_{\min}^2 + d_{\min}^2)/2 - (1+\delta)}}. \tag{A.30}$$

If $(\tilde{d}_{\min}^2 + d_{\min}^2)/2 \leq (1+\delta) + (1-\delta)$ ( $\leq (\tilde{d}_{\max}^2 + d_{\min}^2)/2$ ), we have from Lemma A.3 that $\overline{\eta}_s(d_{\min}) = \overline{E}(d_{\min})$. The associated magnitude triple is $(|\overline{PB}|, |\overline{PC}|, |\overline{BC}|) = (\sqrt{1-\delta} \cdot \xi, \sqrt{1+\delta} \cdot \xi, d_{\min} \cdot \xi)$; by using the law of cosines we have

$$\cos(\angle BPC) = \frac{(1+\delta) + (1-\delta) - d_{\min}^2}{2\sqrt{1+\delta}\sqrt{1-\delta}}. \tag{A.31}$$

Let's turn to consider the other case $(1+\delta) - d_{\min}^2 < 0$. As such, we have $(\cos(\angle BDC))' > 0$ (see (A.26)), and $\cos(\angle BDC)$ is therefore an increasing function of $|\overline{DB}|$. Since $|\overline{DB}| < |\overline{U_{l,s}(d_{\min})B}|$ for all $D \in \mathcal{CV}(P, U_{l,s}(d_{\min})) \setminus \{U_{l,s}(d_{\min})\}$ (again, this can be easily verified based on Figures A.4-(b) and A.4-(c)), the optimal solution is $U_{l,s}(d_{\min})$, and the resultant $\angle BU_{l,s}(d_{\min})C$ is given precisely as in (A.29).

Finally, we note that under the considered assumption $\tilde{d}_{\max}^2 - d_{\min}^2 \geq 0$, we have from (4.2) that $\cos(\angle BDC) \geq 0$ for all $D$. In case that the values of computed $\cos \alpha_{\min}$ are greater than one, we must have $\cos \alpha_{\min} = 1$, and hence $\alpha_{\min} = 0$. $\square$

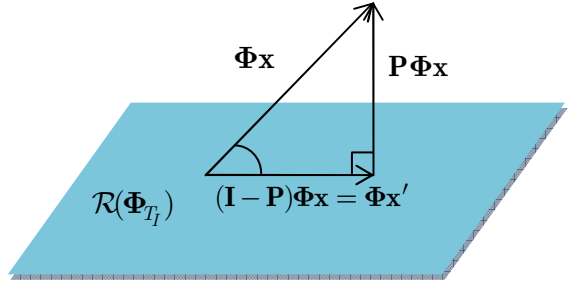

Figure A.11. Schematic description of orthogonal projection of $\mathbf{\Phi x}$ onto the column space of $\mathbf{\Phi}_{T_I}$.

*K. Proof of Theorem 5.1*

Since $\|\mathbf{u} \pm \mathbf{v}\|_2^2 = \|\mathbf{u}\|_2^2 + \|\mathbf{v}\|_2^2 \pm 2\|\mathbf{u}\|_2 \|\mathbf{v}\|_2 \cos\theta = 2 \pm 2\cos\theta$, we have

$$2 - 2|\cos\theta| \leq \|\mathbf{u} \pm \mathbf{v}\|_2^2 \leq 2 + 2|\cos\theta|. \tag{A.32}$$

Based on (A.32) and RIP, it follows that

$$(1-\delta)\{2 - 2|\cos\theta|\} \leq \|\mathbf{\Phi}(\mathbf{u} \pm \mathbf{v})\|_2^2 \leq (1+\delta)\{2 + 2|\cos\theta|\} \tag{A.33}$$

Now,

$$\begin{aligned}|\langle \mathbf{\Phi u}, \mathbf{\Phi v}\rangle| &\stackrel{(a)}{=} \frac{1}{4}\left\{\|\mathbf{\Phi u} + \mathbf{\Phi v}\|_2^2 - \|\mathbf{\Phi u} - \mathbf{\Phi v}\|_2^2\right\} \\ &\stackrel{(b)}{\leq} \frac{1}{4}\{(1+\delta)[2+2|\cos\theta|] - (1-\delta)[2-2|\cos\theta|]\} = \{\delta + |\cos\theta|\},\end{aligned} \tag{A.34}$$

where (a) follows from the polarization identity, and (b) holds from (A.33). Since

$$\|\mathbf{\Phi u}\|_2 \|\mathbf{\Phi v}\|_2 \geq (1-\delta)\|\mathbf{u}\|_2 \|\mathbf{v}\|_2 = 1-\delta, \tag{A.35}$$

inequality (5.2) follows from (A.34) and (A.35), and since $|\cos\alpha| \leq 1$. $\square$

*L. Proof of Theorem 6.1*

The proof can be done by leveraging the geometric property of the orthogonal projection. Specifically, let us decompose $\mathbf{\Phi x}$ into

$$\mathbf{\Phi x} = \mathbf{P}\mathbf{\Phi x} + (\mathbf{I} - \mathbf{P})\mathbf{\Phi x} \tag{A.36}$$

By definition $\mathbf{P} \triangleq \mathbf{I} - \mathbf{\Phi}_{T_I}(\mathbf{\Phi}_{T_I}^*\mathbf{\Phi}_{T_I})^{-1}\mathbf{\Phi}_{T_I}^*$, we immediately have

$$\mathbf{I} - \mathbf{P} = \mathbf{\Phi}_{T_I}(\mathbf{\Phi}_{T_I}^*\mathbf{\Phi}_{T_I})^{-1}\mathbf{\Phi}_{T_I}^*, \tag{A.37}$$

which is the orthogonal projection onto $\mathcal{R}(\mathbf{\Phi}_{T_I})$, namely, the column space of $\mathbf{\Phi}_{T_I}$. As a result, the term

$(\mathbf{I} - \mathbf{P})\mathbf{\Phi}\mathbf{x}$ can be expressed as

$$(\mathbf{I} - \mathbf{P})\mathbf{\Phi}\mathbf{x} = \mathbf{\Phi}_{T_I}\tilde{\mathbf{x}}'_I = \mathbf{\Phi}\mathbf{x}'_I, \tag{A.38}$$

where $\tilde{\mathbf{x}}'_I \in \mathbb{R}^{|T_I|}$, and $\mathbf{x}'_I \in \mathbb{R}^p$ is $|T_I|$-sparse (with support $T_I$) that is obtained by padding $p - |T_I|$ zeros to $\tilde{\mathbf{x}}'_I$. Since $\mathbf{P}$ is an orthogonal projection matrix, it follows that (see Figure A.11)

$$\|\mathbf{P}\mathbf{\Phi}\mathbf{x}\|_2^2 = \|\mathbf{\Phi}\mathbf{x}\|_2^2 \left|\sin \angle(\mathbf{\Phi}\mathbf{x}, \mathbf{\Phi}\mathbf{x}'_I)\right|^2 = \|\mathbf{\Phi}\mathbf{x}\|_2^2 \left(1 - |\cos \angle(\mathbf{\Phi}\mathbf{x}, \mathbf{\Phi}\mathbf{x}'_I)|^2\right). \tag{A.39}$$

Since by assumption $\mathbf{x}$ is a $(K - |T_I|)$-sparse vector, with the aid of (1.2) and (A.38) we then have

$$(1 - |\cos \angle(\mathbf{\Phi}\mathbf{x}, \mathbf{\Phi}\mathbf{x}'_I)|^2)(1 - \delta)\|\mathbf{x}\|_2^2 \leq \|\mathbf{P}\mathbf{\Phi}\mathbf{x}\|_2^2 \leq (1 - |\cos \angle(\mathbf{\Phi}\mathbf{x}, \mathbf{\Phi}\mathbf{x}'_I)|^2)(1 + \delta)\|\mathbf{x}\|_2^2, \tag{A.40}$$

thereby

$$\left(1 - \max\left|\cos \angle(\mathbf{\Phi}\mathbf{x}', \mathbf{\Phi}\mathbf{x})\right|^2\right)(1 - \delta)\|\mathbf{x}\|_2^2 \leq \|\mathbf{P}\mathbf{\Phi}\mathbf{x}\|_2^2 \leq (1 + \delta)\|\mathbf{x}\|_2^2. \tag{A.41}$$

From (5.3), we have

$$\max\left|\cos \angle(\mathbf{\Phi}\mathbf{x}', \mathbf{\Phi}\mathbf{x})\right|^2 = \begin{cases} |\cos(\alpha_{\min})|^2, & \text{if } 0 < \alpha_{\min} < \alpha_{\max} \leq \pi/2, \\ |\cos(\alpha_{\max})|^2, & \text{if } \pi/2 < \alpha_{\min} < \alpha_{\max} < \pi, \\ \max\left\{|\cos(\alpha_{\min})|^2, |\cos(\alpha_{\max})|^2\right\}, & \text{if } 0 < \alpha_{\min} < \pi/2 < \alpha_{\max} < \pi. \end{cases} \tag{A.42}$$

To explicitly specify $\max\left|\cos \angle(\mathbf{\Phi}\mathbf{x}', \mathbf{\Phi}\mathbf{x})\right|^2$, we first note that the supports of $\mathbf{x}$ and $\mathbf{x}'_I$ do not overlap, which implies $\angle(\mathbf{x}, \mathbf{x}'_I) = \pi/2$. Hence, according to (A.42) and Corollary 5.2, it follows

$$\max\left|\cos \angle(\mathbf{\Phi}\mathbf{x}', \mathbf{\Phi}\mathbf{x})\right|^2 = \min\left\{1, \frac{\delta^2}{1 - \delta^2}\right\} \tag{A.43}$$

Combining (A.41) and (A.43) yields (6.3). $\square$

*M. Proof of Theorem 6.3*

The assertion can be obtained by following the proof in [19]. Firstly, the equation (10) in [19, p-4398] shows that the RIC of the effective sensing matrix (the product of orthogonal projection matrix and the original sensing matrix) is given by $\overline{\delta}_a = \delta/(1-\delta)$. Now, by replacing $\overline{\delta}_a$ with the derived $\overline{\delta}$ in Theorem 6.1, and by going through essentially the same procedures as in the proofs of Lemma 3.3 and Corollary 3.1 in [19, p-4398], it can be readily shown that a sufficient condition for recovering a $K$-sparse y via OMP in exactly $K$ iterations is

$$\|\tilde{\mathbf{x}}\|_\infty > 2\overline{\delta}\|\tilde{\mathbf{x}}\|_2. \tag{A.44}$$

Since $\|\tilde{\mathbf{x}}\|_\infty \geq \|\tilde{\mathbf{x}}\|_2/\sqrt{K}$ (the cardinality of the support of $\tilde{\mathbf{x}}$ is less than $K$), a sufficient condition for (A.44) is thus

$$2\overline{\delta} < 1/\sqrt{K}. \tag{A.45}$$

Now by invoking the definition of $\bar{\delta}$ in (6.3), (A.45) can be rearranged as

$$\delta + \frac{\delta^2}{1+\delta} < \frac{1}{2\sqrt{K}}, \tag{A.46}$$

or equivalently,

$$4\sqrt{K}\delta^2 + (2\sqrt{K} - 1)\delta - 1 < 0. \tag{A.47}$$

Based on (A.47), the plausible range of $\delta$ is thus

$$\frac{1 - 2\sqrt{K} - \sqrt{4K + 12\sqrt{K} + 1}}{8\sqrt{K}} < \delta < \frac{1 - 2\sqrt{K} + \sqrt{4K + 12\sqrt{K} + 1}}{8\sqrt{K}}. \tag{A.48}$$

Since

$$\frac{1 - 2\sqrt{K} + \sqrt{4K + 12\sqrt{K} + 1}}{8\sqrt{K}} < 1 \tag{A.49}$$

and $\delta > 0$, the assertion follows from (A.48) and (A.49). □